\documentclass[10pt,aps,prd,floatfix,notitlepage,showpacs,nofootinbib,superscriptaddress,twocolumn,showkeys]{revtex4-1}
\usepackage{amssymb,amsmath}
\usepackage[colorlinks=magenta, colorlinks, linkcolor=blue, citecolor=magenta]{hyperref}
\usepackage{graphicx}

\begin{document}

	
\title{Effective gravitational coupling in modified teleparallel theories}
	
	
\author{Habib Abedi}
\email{h.abedi@ut.ac.ir}
\affiliation{Department of Physics, University of Tehran, North Kargar Avenue, Tehran, Iran.}
	
\author{Salvatore Capozziello}
\email{capozziello@na.infn.it}
\affiliation{Dipartimento di Fisica, Universit\`a di Napoli ''Federico II'', Via Cinthia, I-80126, Napoli, Italy,}
\affiliation{Istituto Nazionale di Fisica Nucleare (INFN), Sez. di Napoli, Via Cinthia, Napoli, Italy,}
\affiliation{Gran Sasso Science Institute, Via F. Crispi 7,  I-67100, L'Aquila, Italy.}

\author{Rocco D'Agostino}
\email{rocco.dagostino@roma2.infn.it}
\affiliation{Dipartimento di Fisica, Universit\`a degli Studi di Roma ``Tor Vergata'', Via della Ricerca Scientifica 1, I-00133, Roma, Italy.}
\affiliation{Istituto Nazionale di Fisica Nucleare (INFN), Sez. di Roma ``Tor Vergata'', Via della Ricerca Scientifica 1, I-00133, Roma, Italy.}

\author{Orlando Luongo}	
\email{orlando.luongo@lnf.infn.it}
\affiliation{Istituto Nazionale di Fisica Nucleare, Laboratori Nazionali di Frascati, 00044 Frascati, Italy.}
\affiliation{School of Science and Technology, University of Camerino, I-62032, Camerino, Italy.}
\affiliation{Department of Mathematics and Applied Mathematics, University of Cape Town, Rondebosch 7701,
Cape Town, South Africa.}
\affiliation{Astrophysics, Cosmology and Gravity Centre (ACGC), University of Cape Town, Rondebosch 7701,
Cape Town, South Africa.}


\begin{abstract}
In the present study, we consider an extended form of teleparallel Lagrangian $f(T,\phi,X)$, as function of a scalar field $\phi$, its kinetic term $X$ and the torsion scalar $T$.
We use linear perturbations to obtain the equation of matter density perturbations on sub-Hubble scales. The gravitational coupling is modified in scalar modes with respect to the one of General Relativity, albeit vector modes decay and do not show any significant effects.
We thus extend these results by involving multiple scalar field models. Further, we study conformal transformations in teleparallel gravity and we obtain the coupling as the scalar field is non-minimally coupled to both torsion and boundary terms.
Finally, we propose the specific model $f(T,\phi,X)=T + \partial_\mu \phi\  \partial^\mu \phi +\xi T \phi^2$. To check its goodness, we employ the observational Hubble data, constraining the coupling constant, $\xi$, through a Monte Carlo technique based on the Metropolis-Hastings algorithm. Hence, fixing $\xi$ to its best-fit value got from our numerical analysis, we calculate the growth rate of matter perturbations and we compare our outcomes with the latest measurements and the predictions of the $\Lambda$CDM model.
\end{abstract}

\date{\today}

\maketitle


\section{Introduction}

The late-time accelerated expansion of the universe has been widely confirmed by low and high-redshift observations \cite{Perlmutter,Tegmark,Allen,Planck}. Although observations show that the cosmic speed up is a consolidate fact, a corresponding physical explanation of this process still remains one of the main challenges of modern cosmology. Several models have been suggested to describe this caveat \cite{orl}, spanning from adding unconventional fluids, passing through modified matter density, up to the introduction of extended theories of gravity (see \cite{LiWang}). Among all models, the \emph{teleparallel} description of gravity has recently reached much attention~\cite{Hayashi,Hayashi2,Aldrovandi,Hehl,Maluf,Abedi3,orl1,orl2,rocco}.

In its simplest description, teleparallel gravity furnishes a gravitational Lagrangian which coincides with the Ricci scalar up to a boundary term. As a consequence of this prescription, teleparallel gravity turns out to be equivalent to the standard Einstein-Hilbert Lagrangian, behaving like General Relativity. In particular, one way to tackle the dark energy problem is  to consider new dynamical degrees of freedom inside the teleparallel scheme. For example, theories in which one replaces the Lagrangian by $f(T)$ functions, i.e. where the torsion scalar  is replaced by a nonlinear function $f(T)$~\cite{Bengochea,Linder,Ferraro,Oikonomou,Ferraro1,Ferraro2}, represent a
viable framework naively inspired by $f(R)$-gravity. This prescription can be even extended by a more general form based on $f(T,B)$ functions in order to include both torsion and the boundary term~\cite{Wright1,Wright2}. This leads to a generalization of both $f(R)$ and $f(T)$ classes of models and have been also studied as $f(R,T)$ in \cite{Myrzakulov}.

In addition, one can use a scalar field responsible for the expansion, providing a \emph{scalar-tensor teleparallel dark energy} acting as solution to the cosmic acceleration
problem~\cite{Panotopoulos,Campo,Geng2,Geng4,Xu,Geng3} and analogous to \cite{Barvinsky}. In this picture, the scalar field should
be non-minimally coupled to gravity. This has been proved by working on renormalization of scalar-tensor theory and in the context of quantum corrections  on curved spacetime~\cite{Sahni,Uzan,Bartolo,Faraoni,Elizalde,Hrycyna,Hrycyna2}.

The most common way is to include a single scalar field, although this choice is not unique. Indeed, even though the use of single field models is consistent with data, the idea to consider multiple field models during inflation and the late-time expansion~\cite{multifield} is still possible. The multi-field can also be non-minimally coupled to gravity.

In the case of General Relativity, using a convenient conformal transformation, one can write the Lagrangian in a particular frame where such a coupling does not appear, i.e. the Einstein frame. In multiple field models, these transformations lead to a curved field space~\cite{Abedi,Kaiser}, i.e. with non-canonical kinetic terms.
Unfortunately, no Einstein frames exist in teleparallel gravity, not even in single field model~\cite{Yang,Wright3}.

\noindent The torsion scalar is not a local Lorentz scalar (i.e. under $h^{A}_{\phantom{A} \mu}=\Lambda^{A}_{\phantom{B} B}(x)\,
h^{B}_{\phantom{B} \mu} $). The violation of local Lorentz symmetry
gives six extra degrees of freedom \cite{Li2}, which disappear in highly symmetric spacetime as in the case of the Friedmann-Lema\^itre-Robertson-Walker (FLRW) metric, or in the simplest model of teleparallel gravity. Structures in the universe are generated from primordial density perturbations when they enter the horizon.
The structure formation is a well-known phenomenon in General Relativity. The linear matter
perturbations on sub-Hubble scales satisfy the following equation~\cite{Ellis}
\begin{equation}
\ddot{\delta}_{\rm m}+2H\dot{\delta}_{\rm m}-4 \pi G\rho_{\rm m}
\delta_{\rm m}=0\ ,
\end{equation}
where $H$ is the Hubble parameter, $\delta_{\rm m}$ is the matter energy density contrast, $G$ is the gravitational coupling and the dot stands for the cosmic time derivative.
Furthermore, observations  indicate the gravitational coupling
$ G $ can be a function of time,
$ \big| \dot{G} / G \big|  <  4.10 \times 10^{-11} {\rm yr^{-1}} $ (see \cite{Ray,Uzan2}).
Using a modified theory of gravity,
the equation of matter perturbations' growth in small scales can be obtained by introducing
an effective gravitational coupling, $G_{ \rm eff}$.
Subsequently, growth of matter perturbations  becomes scale dependent. This can be used to discriminate between proposed models of dark energy.

The aim of this work is to derive the equations of matter density perturbations in generalized teleparallel gravity non-minimally coupled to a scalar field.
In the Sec.~\ref{background spacetime}, we review teleparallel gravity.
In Sec~\ref{f_tele}, we study $f(T,\phi,X)$ theory.
Then, Sec.~\ref{cosmological perturbations} is devoted to  vierbein-based scalar and vector perturbations in linear order around FLRW spacetime. Under a quasi-static approximation, we obtain the effective gravitational coupling deep inside the Hubble radius. Some works have
investigated the effects of teleparallel  extra degrees of freedom on matter perturbations' growth  in a special case of scalar-tensor modification~\cite{Geng3} and
$f(T)$-gravity~\cite{Dent,Chen,Zheng,Wu1,Nesseris}. Our work covers them as specific cases.
It is convenient to study teleparallel extra degrees in a more general form of the Lagrangian that contains the non-minimal coupling between scalar field and scalar torsion.
We  extend our calculation to multiple scalar fields non-minimally coupled to gravity in Sec.~\ref{Multiple scalar field model} (for the General Relativity case, see \cite{Abedi}).
We then study conformal transformations in  Sec.~\ref{Conformal transformation}.
We show that the torsion scalar non-minimally coupled to scalar fields and the boundary term in the form of  $f(-T+B,\phi^i)$  has Einstein frame and respects the local Lorentz symmetry.
In  Sec.~\ref{STB}, we obtain the effective gravitational coupling for scalar field non-minimally coupled to both torsion scalar and the boundary term. This work can be useful to test modified teleparallel gravity in the future observations, e.g. large scales structures, CMB and weak lensing.
In Sec.~\ref{observations}, we focus on a pure teleparallel theory with a vanishing potential. We thus examine the theoretical predictions of such a model by comparing the growth rate of matter perturbations with the outcomes of the concordance $\Lambda$CDM model and the most recent observations.
Finally, we conclude by summarizing our findings in Sec.~\ref{conclusion}.

Throughout this study, $\eta_{AB}=\textrm{diag}(1,-1,-1,-1)$ is used as an orthogonal metric.  Greek indices refer to the coordinates of the manifold, while capital Latin indices refer to the tangent space. We use $ i,j,k...=1,2,3 $ and $a,b,c...=1,2,3 $ for their spatial parts, respectively. We also set $M_{\rm Pl}^2=(8\pi G)^{-1}=1$, where $M_{\rm Pl}$ is the reduced Planck mass.


\section{A review of teleparallel gravity}
\label{background spacetime}

Vierbein fields $h^{A}_{\phantom{A} \mu}$ represent a set of orthonormal basis at each point of the tangent space of the manifold $ ({\cal M},g_{\mu \nu}) $.  The metric tensor on this manifold is related to vierbein fields by
\begin{equation}
g_{\mu \nu}=\eta_{AB} h^{A}_{\phantom{A} \mu} h^{B}_{\phantom{B} \nu}.
\end{equation}
Considering a class of frames in which spin connection is zero, we use the curvatureless Weitzenb\"ock connection as follows
\begin{equation}
\tilde{\Gamma}^{\alpha}_{\phantom{\alpha} \mu\nu}:=h_{A}^{\phantom{A} \alpha}\partial_{\nu}
h^{A}_{\phantom{A} \mu}=- h^{A}_{\phantom{A} \mu}\partial_{\nu} h_{A}^{\phantom{A} \alpha
} .
\end{equation}
This connection is used for parallel transport of vierbein fields, i.e.
$\tilde{\nabla}_\mu h_A^{\phantom{A}\nu}=0$, where $\tilde{\nabla}_\mu$ is covariant derivatives with respect to the Weitzenb\"ock connection.
Using this connection results in the nonvanishing tensor and scalar torsion, respectively,
\begin{align}
T^{\alpha}_{\phantom{\alpha} \mu\nu} &= h_{A}^{\phantom{A} \alpha} \left( \nabla_{\mu} h^{A}_{\phantom{A}\nu} -\nabla_{\nu} h^{A}_{\phantom{A} \mu} \right),
\\
T &= T^{\alpha}_{\phantom{\alpha} \mu\nu} S_{\phantom{\mu\nu} \alpha}^{\mu\nu},
\end{align}
where $ \nabla_{\mu} $ is the covariant derivative defined by the
Levi-Civita connection and
\begin{align}
S^{\alpha\mu\nu}=&\frac{1}{4} \left(T^{\alpha\mu\nu}+T^{\mu\alpha\nu}-T^{\nu\alpha\mu} \right)\nonumber \\&-\frac{1}{2}\left(g^{\alpha
	\nu}T^{\beta\mu}_{\phantom{\beta \mu} \beta}-g^{\mu \alpha}T^{\beta \nu}_{\phantom{\beta \nu} \beta} \right)
\end{align}
is the superpotential tensor, which is antisymmetric under last two indices.
In the Riemannian geometry, curvature does not vanish differently from the torsion which is zero. In this picture $T^\alpha_{\phantom{\alpha}\mu \nu}$ represents a tensor with no geometric meaning. However, we use Weitzenb\"ock connection where $T^\alpha_{\phantom{\alpha}\mu \nu}$ is interpreted as a torsion tensor.
The metric is invariant under local Lorentz transformations in the tangent space. In this case, the vierbein fields becomes
\begin{equation}
h^A_{\phantom{A} \mu} \rightarrow \Lambda^A_{\phantom{A} B}(x^\nu) \, h^B_{\phantom{B} \mu}.
\end{equation}
Consequently, the torsion tensor transforms as
\begin{equation}
T^\alpha_{\phantom{\alpha} \mu \nu} \rightarrow T^\alpha_{\phantom{\alpha} \mu \nu} + \Lambda^{\phantom{B} A}_B h^{\phantom{A} \alpha}_{A} \left(h^C_{\phantom{C} \nu} \, \partial_\mu  -
h^C_{\phantom{C} \mu} \, \partial_\nu
\right) \Lambda^B_{\phantom{B} C}.
\end{equation}
The torsion scalar can be written as $T=-R+B$, where  $ B=2 \nabla_\nu T^{\mu \phantom{\mu} \nu}_{\phantom{\mu} \mu} $ is  the boundary term, and $R$ is Ricci scalar constructed from Levi-Civita connection.


\section{$f(T,\phi,X)$ theory}
\label{f_tele}

The general form of an action, $S$, containing the torsion scalar $T$, a scalar field $ \phi $ and the kinetic term $X\equiv\nabla_\mu \phi \, \nabla^\mu \phi$, is given by~\cite{Abedi2}
\begin{equation}
S= \int {\rm d}^4 x\, h \left[\frac{1}{2} f(T,\phi,X) +\mathcal{L}_{\rm m} \right] , \label{action1}
\end{equation}
where $h$ is determinant of vierbein fields $h^{A}_{\phantom{A} \mu}$, and ${\cal L}_{\rm m}$ is matter Lagrangian.
Varying the action with respect to vierbein fields yields to~\eqref{action1}
\begin{align}
\Theta ^{\phantom{A} \mu}_{A}=&\frac{1}{2}f h_{A}^{\phantom{A} \mu}+ f_{T} \left[ h^{-1}\partial_{\nu} \left( hh^{\phantom{A}
	\rho}_{A}S^{\phantom{\rho} \mu \nu}_{\rho} \right)-h^{\phantom{A} \gamma}_{A}S^{\rho \beta
	\mu}T_{\rho \beta \gamma} \right]
\nonumber \\	&+ h^{\phantom{A} \rho}_{A}S^{\phantom{\rho}
	\nu \mu}_{\rho} \partial_{\nu} f_T
- \frac{1}{2} f_{X}
h_{A}^{\phantom{A} \nu} \, \partial^{\mu} \phi \,
\partial_{\nu} \phi , \label{field1}
\end{align}
where $ f_T:=\partial f/\partial T $, $ f_X:=\partial f/\partial X$ and $\Theta ^{\phantom{\nu}
	\mu}_{\nu}=h^{A}_{\phantom{A} \nu} \Theta^{\phantom{A} \mu}_{A}=-h^{A}_{\phantom{A}
	\nu}\, \delta \mathcal{L}_{\rm m}/\delta h^{A}_{\phantom{A} \mu} $ is  the matter
energy-momentum tensor. We have dropped the explicit dependence on $f$. Further, variation of the action~\eqref{action1} with respect to scalar field provides the following field equation:
\begin{equation}
\square \phi +\frac{\partial_\alpha f_X}{f_X}\, \partial^\alpha \phi -\frac{f_\phi}{f_X}=0 \label{field2}\ ,
\end{equation}
where $\square \phi=\partial_{\mu}\big(\sqrt{-g}\partial^{\mu}\phi \big) / \sqrt{-g}$.
We can thus write Eq.~\eqref{field1} under a  covariant representation by
\begin{align}
\Theta^{\phantom{\alpha} \mu}_{\alpha}=&f_T G^{\phantom{\alpha} \mu}_{\alpha}+ \frac{1}{2} \delta^{\mu}_{\alpha} (f-f_T
T) + S^{\phantom{\alpha} \nu \mu}_{\alpha}  \partial_{\nu} f_T \nonumber \\ &-\frac{1}{2} f_{X}\; \partial^{\mu} \phi \;
\partial_{\alpha} \phi, \label{field8}
\end{align}
where $G_\alpha^{\phantom{\alpha}\mu}$ is the Einstein tensor.
One requests that the energy-momentum tensor is invariant under local Lorentz
transformation and symmetric. As a consequence, its antisymmetric part turns out to
vanish:
\begin{equation}
\left( S_{\alpha}^{\phantom{\alpha} \lambda\nu}g^{\alpha\mu}-S_{\alpha}^{\phantom{\alpha} \lambda\mu
}g^{\alpha\nu} \right)  \partial_{\lambda} f_T =0 \label{cons1}.
\end{equation}
According to the cosmological principle and recent observations, we choose the spatially flat FLRW as the background spacetime,
\begin{equation}
{\rm d}s^2 = {\rm d}t^2 - a^2 (t) \, \delta_{ij}\, {\rm d}x^i\, {\rm d}x^j,
\end{equation}
where $a(t)$ is the scale factor. Under the trivial choice of vierbein fields,
\begin{equation}
\left(  h_{\phantom{A} \mu}^{A} \right) = {\rm diag}\left(1,a,a,a\right)\ ,
\end{equation}
the field equations read
\begin{align}
\rho_{\rm m}& =3H^2 f_T-\frac{1}{2} (f-f_T T)+\frac{1}{2} f_{X} \dot{\phi}^{2}  , \label{edensity}
\\
p_{\rm m}&=- \big(3H^2 + 2 \dot{H} \big) f_T+\frac{1}{2} (f-f_T T) -H \, \partial_{0} f_T, \label{pressure}
\\
0&=\ddot{\phi}+\left( 3H+\frac{\partial_0 f_X}{f_X}\right)\dot{\phi}-\frac{f_\phi}{f_X},
\end{align}
where we considered that $\phi$ depends only on time at the background level. The quantities $ \rho_{\rm m} $ and $p_{\rm m}$ are the energy density and pressure of matter respectively. We can rewrite Eqs.~\eqref{edensity} and \eqref{pressure} as follows
\begin{align}
\rho_{\rm DE}+\rho_{\rm m}& = 3H^2,
\nonumber \\
p_{\rm DE}+p_{\rm m}&= -\big( 3H^2+2\dot{H} \big),
\end{align}
where $\rho_{\rm DE}$ and $p_{\rm DE}$ are defined by
\begin{align}
&\rho_{\rm DE}= 3H^2 (1-f_T)+\frac{1}{2} (f-f_T T) - \frac{1}{2}f_X \dot{\phi}^2, \nonumber \\
&p_{\rm DE}= -(3H^2+2\dot{H}) (1-f_T)-\frac{1}{2} (f-f_T T)+H\, \partial_0 f_T.
\end{align}
Finally, we can now write dark energy equation of state $\omega_{\rm DE}:=p_{\rm DE}/\rho_{\rm DE}$, with  $p_{\rm m}=\omega \rho_{\rm m}$:
\begin{equation}
\omega_{\rm DE}=-\frac{1+\dfrac{2\dot{H}}{3H^2}+ \dfrac{\omega \rho_{\rm m}}{3H^2}}{1-\dfrac{\rho_{\rm m}}{3H^2}}.
\end{equation}


\subsection{Cosmological perturbations}
\label{cosmological perturbations}

The cosmological principle predicts homogeneity and isotropy on large scales. In the FLRW model the complex distribution of galaxies and clusters of galaxies can be described by adopting perturbations. In particular, perturbations as {\it seeds of all structures} can explain the existence of the observed inhomogeneity.

In this section, we discuss these perturbations in vierbein approach around the
homogeneous and isotropic model at sub-Hubble scales.
The vierbein fields for FLRW perturbations can be written as
\begin{align}
h^{(0)}_{\phantom{(0)}\mu}=&\delta^{0}_{\mu}(1+\psi)+a\delta^{i}_{\mu} \, \partial_{i}(F+\alpha)
+a\delta^{i}_{\mu} \left(G_i +\alpha _i \right),
\nonumber \\
h^{(a)}_{\phantom{(a)}\mu}=&a \delta^{a}_{\mu}(1-\varphi)+a
\delta^{i}_{\mu} \left(\partial_{i} \partial^{a} B
+\partial^{a}C_{i}+h^{\; a}_{i} \right)
\nonumber \\ 	  &+a\delta^{i}_{\mu} B^{\phantom{i}a}_{i}
+\delta^{0}_{\mu} \left(\partial^{a}\alpha +\alpha ^a \right),
\end{align}
where $ \psi $, $ \varphi $, $ F $, $ \alpha $ and $ B $ are
scalar quantities; $ G_{i} $, $ \alpha_{i} $ and $ C_{i} $ are vectors; $
h_{ij} $ is a tensor and $ B_{ij} $ with $ B_{ij}+B_{ji}=0 $ has both
scalar and vector parts. Here, all vectors are transverse,
and the tensor is both transverse and traceless.

Perturbations   $ \alpha $, $
\alpha_{i} $ and $ B_{ij} $ do not appear in quantities evaluated from the metric. If
perturbations are small, one can choose a
coordinate system in which vierbein perturbations become small too.
Moreover, the action is invariant under gauge transformations, so that by performing the gauge transformation $ x^{\mu} \rightarrow
\tilde{x}^{\mu}=x^{\mu}+\xi^{\mu} $, the vierbein fields transform as
\begin{equation}
h^{A}_{\phantom{A} \tilde{\mu}}(\tilde{x})=\frac{\partial
	x^{\nu}}{\partial \tilde{x}^{\mu}} h^{A}_{\phantom{A} \nu}(x). \label{G1}
\end{equation}
Using transformation~\eqref{G1},  vierbein perturbations in new gauge become
\begin{align}
\tilde{ h}_{ij}=&h_{ij}, \nonumber \\
\tilde{B}_{ij}=& B_{ij}+\xi_{i,j}^{(v)}/a-\xi_{j,i}^{(v)}/a, \nonumber \\
\tilde{G_{i}}=&G_{i}+\dot{\xi}^{(v)}_i ,\nonumber \\
\tilde{ \alpha}_{i}=&\alpha_{i}-\dot{\xi}_{i}^{(v)}, \nonumber \\
\tilde{C}_{i}=& C_{i}- \xi_{i}^{(v)}/a, \nonumber \\
\tilde{\psi}=&\psi-\dot{\xi}^{0} ,\nonumber \\
\tilde{F}=&  F+\dot{\xi}/a- \xi^{0}/a ,\nonumber \\
\tilde{\alpha}=&\alpha-\dot{\xi} ,\nonumber \\
\tilde{B}=&B-\xi/a ,\nonumber \\
\tilde{\varphi}=&\varphi +H \xi^{0}.
\end{align}
We have used $\xi_{i}:=\big(\xi_{i}^{(v)}+\xi_{,i}\big)/a $, where
$\xi_{i}^{(v)} $ is the transverse part of $ \xi_{i} $. The scalar, vector
and tensor perturbations are not coupled to each other at linear order. Thus, they can be
handled independently.

We now study scalar and vector perturbations in sub-Hubble scales separately.

\subsubsection{Scalar Perturbations}

In this subsection, we consider scalar perturbations, which are related to the density perturbations. By choosing
a convenient gauge, i.e. $ \xi $, $ \xi_{0} $ and $ \xi^{(v)}_{i} $, one
sets $ B $, $ F $ and $ C_{i} $ to zero. Further, proceeding with conformal
Newtonian or zero-shear gauge, the corresponding line element is given by
\begin{equation}
{\rm d}s^2=(1+2\psi) \, {\rm d}t^2-a^2(1-2\varphi)\delta_{ij} \, {\rm d}x^{i} \, {\rm d}x^{j}. \label{line1}
\end{equation}
The metric perturbation $ \psi $ is the generalized Newtonian potential
and $\varphi $ is related to the perturbation in the surface three-curvature. Vierbein fields corresponding to the metric~\eqref{line1} take the following forms:
\begin{align}
h^{(0)}_{\phantom{(0)}\mu}=&\delta^{0}_{\mu}(1+\psi)+a\delta^{i}_{\mu} \, \partial_{i}\alpha ,
\nonumber \\
h^{(a)}_{\phantom{(a)}\mu}=&a \delta^{a}_{\mu}(1-\varphi)+a\delta^{i}_{\mu}
B^{\phantom{i}a}_{i} +\delta^{0}_{\mu} \, \partial^{a}\alpha ,
\end{align}
where $\alpha $ and $ B^{\phantom{i}a}_{i} $ are extra scalar teleparallel degrees and
$\partial_{i}\partial_{j}B^{ij}=0$. The scalar field can be decomposed as
\begin{equation}
\phi(x^{i},t)=\phi(t)+\delta\phi(x^i,t),
\end{equation}
where $ \phi(t) $ and $ \delta\phi(x^i,t) $ are the background and first order perturbations  respectively.
We thus assume that the background matter is made by a perfect fluid source. Up to the first
order of perturbations, the matter energy-momentum becomes
\begin{equation}
\left(  \Theta_{\mu}^{\phantom{\mu} \nu} \right) =
\left(
\begin{array}{cccc}
-(\rho_{\rm m} +\delta\rho_{\rm m}) & -a^{-2}(\rho_{\rm m}+ p_{\rm m}) \, \partial_{i}\delta u \\(\rho_{\rm m} + p_{\rm m}) \, \partial_{i}\delta u & \delta_{j} ^{i} (p_{\rm m}+\delta p_{\rm m})       \end{array}
\right)    ,
\end{equation}
where $\delta u$ is the potential of velocity perturbation. The quantities $ \delta \rho_{\rm m} $ and $ \delta p_{\rm m} $ are matter perturbations for the density and pressure respectively.
By means of Eq.~\eqref{field8}, the first order of
perturbation in energy density becomes
\begin{align}
- \delta \rho_{\rm m} =& -3H^2 \, \delta f_T +f_T \left[ 6H(\dot{\varphi}+H\psi)+\frac{2}{a^2}\partial^2 \varphi \right]
\nonumber \\ &+\frac{1}{2} \big(\delta f -\delta f_T \, T-f_T \, \delta T\big) -X \, \delta f_X
\nonumber \\ &- f_X \dot{\phi} \, \partial_0 \delta \phi  , \label{energy}
\end{align}
where $ \partial^{2}:=\delta^{ij}\partial_{i} \partial_{j}$  and
the momentum density is
\begin{align}
(p_{\rm m}+\rho_{\rm m}) \,\partial_{i} \delta u =& -2 f_T\, \partial_{i}
(\dot{\varphi} +H \psi ) +H \,\partial_i \delta f_T
\nonumber \\	&+ \frac{1}{2} f_X \, \partial_0 \phi \,  \partial_i \delta \phi . \label{25}
\end{align}
It is worth remarking that  since  background quantities
have not spatial dependence, spatial derivatives can be dropped in Eq.~\eqref{25}. Next, the energy flux is obtained as
\begin{align}
-(p_{\rm m}+\rho_{\rm m})\,\partial^{i} \delta u =&2 f_T \,\partial^{i}
(\dot{\varphi} +H \psi )
\nonumber \\	&+\Big(\partial^i \varphi
+ \frac{1}{2} \partial_j B^{ij} \Big)\partial_0 f_T
\nonumber \\	&	-\frac{1}{2} f_X \,\partial_0 \phi \, \partial^i \delta \phi .
\label{flux}
\end{align}
We can also write the diagonal components of momentum flux as follows
\begin{align}
\delta p_{\rm m}=&-\delta f_T \left(2\dot{H}+3H^2\right)
\nonumber \\
&	+f_T  \Bigg[ 6H (\dot{\varphi}+H\psi)+2 \left(\ddot{\varphi}+H\dot{\psi}+2\dot{H}\psi \right) \Bigg]
\nonumber \\
&	+\frac{1}{2} \left(\delta f-\delta f_T -f_T \delta T\right)
+\left(\dot{\varphi}+2H\psi\right)\,\partial_0 f_T
\nonumber \\
&	-H\,\partial_0 \delta f_T .
\label{diagonal}
\end{align}
Its anisotropic part becomes
\begin{equation}
f_T \partial^{i} \partial_{j} (\psi - \varphi)+a \, \partial^{i} \partial_{j} \alpha \, \partial_0 f_T =0. \label{G2}
\end{equation}
The relation~\eqref{G2} can be written as
\begin{equation}
\psi - \varphi = - \frac{\partial_0 f_T}{f_T} a \alpha \label{aniso}.
\end{equation}
In General Relativity or the simplest model of teleparallel gravity, when $f_T=1$, the anisotropic part of field equations~\eqref{field1} vanishes, i.e. $ \psi=\varphi $.
Instead, Eq.~\eqref{aniso} shows one of the important consequences of using modified teleparallel
gravity, i.e. $\psi \neq \varphi$. If we neglect
the extra degree $ \alpha $ in Eq.~\eqref{aniso}, the anisotropic stress will vanish.
In $f(R,\phi,X)$-theory the extra degrees do not appear. The anisotropic part, however, does not  vanish and it is proportional to $\delta f_R$ \cite{Abedi}.
Constraint ~\eqref{cons1} is automatically satisfied at the background level. However, at linear
order of perturbations, it becomes
\begin{equation}
\left(\partial^i \varphi +\frac{1}{2}\partial_j B^{ij} \right)\, \partial_0 f_T +H\, \partial^i \delta f_T=0 . \label{C1}
\end{equation}
By using $
\partial_{i} \partial_{j}B^{ij}=0$, Eq.~\eqref{C1} assumes the following form
\begin{equation}
H\partial_{i} \partial^{i}  \delta f_T +\partial_0 f_T \partial_{i}
\partial ^{i} \varphi =0.
\end{equation}
We consequently have
\begin{equation}
\varphi =-\frac{H}{\partial_0 f_T} \delta f_T \label{29}.
\end{equation}
The scalar field's equation~\eqref{field2} at linear order of perturbations can be written as
\begin{align}
0=&\delta f_X \left(\ddot{\phi}-3H\dot{\phi} \right) +\partial_0 \delta f_X \, \dot{\phi} + \partial_0 f_X \partial_0 \delta \phi -\delta f_{\phi} \nonumber \\
&+f_X \Big\{ -2\psi \ddot{\phi}
-6H\varphi \dot{\phi}
+\partial_0^2 \delta \phi
- \frac{1}{a^2} \partial^2  \delta \phi
\nonumber \\
&\qquad	-\dot{\psi} \dot{\phi} + \left[ 6H(\psi+\varphi) -3 \dot{\varphi} \right] \dot{\phi}
+3H \partial_0 \delta \phi \Big\} , \label{field3}
\end{align}
Torsion scalar up to linear order is $T=-6H^2+12 (\dot{\varphi}+H\psi)+4\partial^2 \zeta / a^2$ where $\zeta:=aH\alpha $ is a dimensionless quantity.
In sub-Hubble scales ($ k\gg aH $)  with the condition $ \dot{Q} \leq |HQ|$ (where $ Q= \phi, \varphi, \psi $), Eq.~\eqref{field3} takes the following form
\begin{equation}
-\frac{f_X}{a^2}\partial^{2}\delta\phi -\delta f_\phi=0. \label{field5}
\end{equation}
The linear order perturbation of $f_\phi$ can be calculated as
\begin{equation}
\delta f_\phi\simeq f_{\phi T} \delta T + f_{\phi \phi} \delta \phi \simeq f_{\phi T} \dfrac{4}{a^2}\partial^2 \zeta +f_{\phi \phi} \delta \phi  .
\label{30}
\end{equation}
If we insert the above expression into Eq.~\eqref{field5}, we obtain
\begin{equation}
-\frac{f_X}{a^2}\partial^{2}\delta\phi-f_{\phi \phi} \delta \phi\simeq  f_{\phi T} \dfrac{4}{a^2}\partial^2 \zeta 	\ ,	
\end{equation}
which yields
\begin{equation}
\delta\phi\simeq -4\dfrac{(k/a)^2 f_{\phi T}\ \zeta}{(k/a)^2 f_X-f_{\phi\phi}}\ .
\label{29bis}
\end{equation}
Considering that field's derivatives do not appear in $f_T$ and using  Eq.~\eqref{30} and Eq.~\eqref{29bis}, from Eq.~\eqref{29} we get
\begin{equation}
\varphi =\frac{4H}{\partial_0 f_T} \left[ \frac{f^2_{\phi T}}{(k/a)^2 f_X-f_{\phi\phi}}+f_{TT} \right]\left(\dfrac{k}{a}\right)^2 \zeta \ . \label{ab1}
\end{equation}
Therefore, by using Eqs.~\eqref{aniso} and \eqref{ab1}, we obtain the strength of anisotropic stress:
\begin{align}
\eta:=&\frac{\psi-\varphi}{\varphi}
\nonumber \\
=&-\frac{(\partial_0 f_T)^2}{4H^2f_T} \left[ \frac{f^2_{\phi T}}{(k/a)^2 f_X-f_{\phi\phi}}+f_{TT} \right]^{-1}\left(\dfrac{k}{a}\right)^{-2}. \label{grav1}
\end{align}
Using energy density~\eqref{energy}, the
Poisson equation becomes
\begin{equation}
\left( \frac{k}{a} \right)^2 \psi =\frac{1}{f_T}
(1+\eta)\frac{\delta\rho_{\rm m}}{2}.
\end{equation}
Hence,  we can obtain the generalized gravitational coupling as follows
\begin{equation}
G_{ \rm eff}=\frac{1+\eta}{f_T}. \label{G13}
\end{equation}
We can also find
\begin{equation}
\left( \frac{k}{a} \right)^2 \varphi =\frac{1}{f_T}
\frac{\delta\rho_{\rm m}}{2}\label{grav2}.
\end{equation}
The effective potential $ \Phi_{ \rm eff}:= \varphi+ \psi $, which appears
in the frameworks of Weak Lensing and Integrated Sachs-Wolfe effect, is:
\begin{equation}
\Phi_{ \rm eff}= \frac{2+\eta}{1+\eta} \psi ,
\end{equation}
which is affected by $\eta$.
In the case of pressureless matter, the conservation of the energy-momentum tensor gives
\begin{align}
\dot{	\delta \rho}_{\rm m} + 3 H \; \delta \rho_{\rm m} =& \rho_{\rm m} \left(\frac{k}{a} \right)^2 \delta u +3 \rho_{\rm m} \dot{\varphi}  , \label{G11} \\
\dot{\delta u} =& - \psi . \label{G12}
\end{align}
Then, defining the gauge-invariant matter density contrast as
\begin{equation}
\delta_{\rm m}:=\frac{\delta \rho_{\rm m}}{\rho_{\rm m}} -3H \; \delta u ,
\end{equation}
using Eq.~\eqref{G12} and time derivative of Eq.~\eqref{G11},
deep inside the Hubble radius, one gets
\begin{equation}
\ddot{\delta}_{\rm m}+ 2 H \dot{\delta}_{\rm m} - \left( \frac{k}{a} \right)^2 \psi \simeq 0\ ,
\end{equation}
and, finally,
\begin{equation}
\ddot{\delta}_{\rm m}+2H\dot{\delta}_{\rm m}-G_{ \rm eff} \, \rho_{\rm m}\frac{\delta_{\rm m}}{2} \simeq 0. \label{growth}
\end{equation}
We can use Eq.~\eqref{G13} to write the gravitational coupling in some models, e.g.:

\begin{itemize}
	\item
	Using  $ f=T+ 2 P(\phi, X) $, where $ P(\phi, X) $ is an arbitrary function of the scalar field and the kinetic term, we obtain the standard gravitational coupling, i.e. $ G_{\rm eff}=1 $.
	\item
	In the case of $f=T+F(T)$, where $f$ is an arbitrary function of the torsion scalar, the effective gravitational coupling reduces to
	\begin{equation}
	G_{ \rm eff}\simeq \frac{1}{1+F_T(T)},
	\end{equation}
	which is the same as in \cite{Zheng}.
	\item The scalar field non-minimally coupled  to the torsion scalar by
	$f=F(\phi)\, T-\partial_\mu \phi \, \partial^\mu \phi -2U(\phi)$  leads to the following effective gravitational coupling:
	\begin{equation}
	G_{ \rm eff}\simeq \frac{1}{F(\phi)} \left(1-\frac{\dot{\phi}^2}{2H^2 F(\phi)} \right).
	\end{equation}
	If we consider $F(\phi)= 1+\xi \phi^2$, the above formula agrees with \cite{Geng3}.
	\item Inspired from the Brans-Dicke theory \cite{Brans}, we consider
	$ f=\phi T+2\omega_{\rm BD} X / \phi $, where $\omega_{\rm BD}$ is a constant. In this case, the gravitational coupling takes the form
	\begin{equation}
	G_{\rm eff} \simeq \frac{1}{\phi}\left(1-\frac{\omega_{\rm BD} \, \dot{\phi}^2}{2H^2\phi^2}\right). \label{BD}
	\end{equation}
	We can see	$ \omega_{\rm BD} \rightarrow 0  $ results in $ G_{\rm eff} \rightarrow 1/\phi$.
\end{itemize}

\subsubsection{Vector Perturbations}

In this paragraph, we study the vector perturbations.
Vector parts of vierbein perturbations are given by
\begin{align}
h^{(0)}_{\phantom{(0)}\mu}=&\delta^{0}_{\mu}+a\delta^{i}_{\mu} \left(G_i +\alpha
_i\right),
\nonumber \\
h^{(a)}_{\phantom{(0)}\mu}=&a\delta^{a}_{\mu}+a\delta^{i}_{\mu}
\left( \partial^{a} C_i +B^{\phantom{i}a}_{i} \right) +\delta^{0}_{\mu}\alpha ^a ,
\end{align}
where $ B^{\phantom{i}a}_{i} $ and $ \alpha _i $ have four extra vector degrees
of freedom and $\partial^{i}\alpha_{i}=0 $. The scalar field $\phi$
has only scalar perturbations and consequently does not contribute to vector
perturbations. The vector part of the field equation~\eqref{field1} gives the
momentum density perturbation as
\begin{equation}
\delta \Theta_{i}^{0}=f_T \, \delta G_{i}^{0} ,
\end{equation}
or, equivalently, one can write
\begin{equation}
a \rho_{\rm m} \, \delta u^{V}_{i}=\frac{1}{2} f_T\, \partial^{2} G_{i}  .
\end{equation}
The right hand side of the above relation becomes proportional to $ a^{-2} $. The
vector perturbations of Eq.~\eqref{cons1}, for $
\mu=0 $ and $ \nu=i $, become
\begin{equation}
\beta^{i}=3aHG^{i}\ .
\end{equation}
For $ \mu=i $ and $ \nu=j $, we get
\begin{equation}
\partial^{i} \left(G^{j}+\alpha^{j} \right)=\partial^{j} \left(G^{i}+\alpha^{i}\right) .
\end{equation}
Therefore, $ \alpha^{i} $ and $ \beta^{i} $ decay as $ a^{-2}$ and
consequently have no significant effect in cosmological
evolution. This behaviour is the same as $ f(T) $-gravity~\cite{Wu1}.
The constraints given here and in \cite{Wu1} are modified by replacing $ f_{T}(T) $ by $ f_T(T,\phi,X) $.
Since these constraints are time-dependent background quantities, they lead to the same equations.

\subsubsection{Super-horizon scales}

In the previous paragraphs, we investigated the scalar perturbations on sub-Hubble scales. However, it has been shown in \cite{Barrow} that deviations in $f(T)$ gravity from the $\Lambda$CDM model become important on super-horizon scales $(k\ll aH)$, at which the evolution of the perturbations results very different.
Also, in \cite{Wu1} it was found that significant deviations from General Relativity arise from extra degrees of freedom in modified teleparallel gravity on super-horizon scales. We expect the same behaviour in our picture. In particular, the combination of Eqs.~\eqref{energy}, \eqref{flux} and \eqref{diagonal} evaluated in the super-Hubble limit would provide $\delta_m$ as sum of a constant plus a term that encodes the departure from General Relativity due to the new degrees of freedom.


\section{Multiple scalar field model}
\label{Multiple scalar field model}

In this section, we extend the action~\eqref{action1} to multiple scalar fields $f(T,\phi^I,X)$, where non-canonical kinetic terms and metric of field space are $2 X:= \mathcal{ G}_{IJ} (\phi^K) \nabla_\mu \phi^I \nabla^\mu \phi^J$ and $ {\cal G}_{IJ}(\phi^K) $ respectively.
Indices $I,J,K,...=1,2,3,..,N$ stand for $N$ scalar fields. Variation of action with respect to vierbein fields, whereas scalar fields yield respectively,
\begin{align}
f_T G^{\phantom{\alpha}\mu}_\alpha+\frac{1}{2} \delta^\mu_\alpha (f-f_T T)+S_\alpha^{\phantom{\alpha} \nu \mu} \, \partial_\nu f_T
&	\nonumber \\ - \frac{1}{2} f_X {\cal G}_{IJ} \, \nabla_\alpha \phi^I \, \nabla^\mu \phi^J  =&
\Theta^\mu_\alpha  , \label{field6}
\\
\nabla_\mu \nabla^\mu \phi^I+ \Gamma^I_{\phantom{I}JK} \, \nabla_\mu \phi^J \, \nabla^\mu \phi^K+\frac{\partial_\mu f_X}{f_X} \nabla^\mu \phi^I
&\nonumber \\
-{\cal G}^{IJ}\frac{f_J}{f_X} =&0 , \label{field7}
\end{align}
where $\Gamma^I_{\phantom{I}JK}$ is the Levi-Civita symbol constructed from field space metric ${\cal G}_{IJ}$, and $f_I$ denotes derivative of $f$ with respect to the scalar field $\phi^I$. Eq.~\eqref{field6}  in FLRW background gets the following form:
\begin{align}
\rho_{\rm m} & = 3H^2 f_T-\frac{1}{2} (f-f_T T)+\frac{1}{2} f_{X} {\cal G}_{IJ} \dot{\phi}^I \dot{\phi}^J ,
\\
p_{\rm m} &= - \big(3H^2 + 2 \dot{H} \big) f_T+\frac{1}{2} (f-f_T T) -H \partial_{0} f_T .
\end{align}
Field equation~\eqref{field7} also takes the following compact form in the background:
\begin{equation}
{\cal D}_t \left( f_X a^3 \dot{\phi}^I \right)=f^Ia^3,
\end{equation}
where $ {\cal D}_t  $  defined by
\begin{equation}
{\cal D}_t  \dot{\phi}^I:=\ddot{\phi}^I+\Gamma^I_{\phantom{I}JK} \dot{\phi}^J \dot{\phi}^K
\end{equation}
acts as ordinary derivatives on  quantities that have not capital Latin indices.
On sub-Hubble scales and in the quasi-static approximation, only Eq.~\eqref{field5} changes,
\begin{equation}
-f_X \frac{\partial^2}{a^2} \delta \phi^I -f_J \, {\cal G}^{IJ}_{\phantom{IJ} ,K} \, \delta \phi^K - {\cal G}^{IJ} \, \delta  f_J=0.
\end{equation}
With straightforward calculation we arrive to the following strength of anisotropic stress:
\begin{equation}
\eta_{\rm eff}:=\frac{(\partial_0 f_T)^2}{H^2 f_T} \left[ f_{TI}(\beta^{-1} \gamma)^{I} - 4 f_{TT} \left(\frac{k}{a}\right)^2 \right]^{-1}  ,
\label{grav3}
\end{equation}
where $ \beta $ and $\gamma$ are background matrices defined as
\begin{align}
\beta_J^{\phantom{I} I}:=& f_X \delta_J^I  \left(\frac{k}{a}\right)^2 - {\cal G}^{IK}_{\phantom{IK} ,J} f_K-\mathcal{G}^{IL}f_{LJ} , \\
\gamma^I :=& -4 {\cal G}^{IJ} f_{JT} \left(\frac{k}{a}\right)^2.
\end{align}
Thus, the effective gravitational coupling can finally be written as follows
\begin{equation}
G_{\rm eff} \simeq  \frac{ 1+\eta_{ \rm eff}}{f_T} \ .
\end{equation}
It is worth noting that, in the case of a single field, Eq.~\eqref{grav3} reduces to Eq.~\eqref{grav1}.

As an example, one can consider the action of $N$ canonical scalar fields non-minimally coupled to the torsion scalar:
\begin{equation}
S=\int {\rm d}^4x \, h \left[ F(\phi^I) \, T - \frac{1}{2} \delta_{IJ} \, \partial_\mu \phi^I \, \partial^\mu \phi^J - V(\phi^I) \right].
\end{equation}
In that case, one has
\begin{equation}
G_{\rm eff} \simeq \frac{1}{F} \left(1-\frac{\dot{F}^2}{2FH^2 \sum_I F_I^{\, 2} }\right).
\end{equation}


\section{Conformal transformations}
\label{Conformal transformation}

So far we have considered scalar fields non-minimally coupled to gravity.
Since the form of field equations are complicated in presence of non-minimal coupling, it is a common procedure to use conformal transformations and to write the equations in the Einstein frame. Shifting from the Jordan frame to the Einstein frame would simplify the equations. As a consequence, it is worth exploring the existence of the Einstein frame in teleparallel gravity.
To do that, we consider the following transformation:
\begin{align}
\hat{h}^A_{\phantom{A} \mu} =& \Omega(x^\alpha) \,  h^A_{\phantom{A} \mu},&
 \hat{h}_A^{\phantom{A} \mu} =& \Omega^{-1}(x^\alpha) \,  h_A^{\phantom{A} \mu} , \label{G6}
\end{align}
where we have used the `hat' notation for quantities in the new frame and $\Omega$ as the conformal factor.
The conformal transformation~\eqref{G6} in FLRW leads to
\begin{align}
{\rm d}\hat{t} =& \Omega(t) \, {\rm d}t,&
 \hat{a} (\hat{t}) &= \Omega(t) \, a(t),&
{\rm d}\hat{x}^i =& {\rm d}x^i.
\end{align}
The vector and tensor perturbations are conformally invariant.
Considering the conformal factor up to linear order of perturbations,
$ \Omega(t,{\bf x})=\Omega(t) \, \left[ 1+\delta \Omega(t,{\bf x}) / \Omega(t) \right] $, one can write
\begin{equation}
\delta \hat{h}_{\mu \nu} = \Omega \,\delta h_{\mu \nu}+ h_{\mu \nu}\, \delta \Omega.
\end{equation}
The perturbation in  the component of vierbein that vanishes in the FLRW background, are conformal invariant.
Comparing  change of vierbein fields under conformal transformation in first order of perturbations, we have
\begin{align}
\hat{\psi} =& \psi + \frac{\delta \Omega}{\Omega},
\nonumber \\
\hat{\varphi} =& \varphi - \frac{\delta \Omega}{\Omega}.
\end{align}
The other scalar perturbations and the extra degrees of freedom are conformal invariant.
Newtonian gauge, considered in this work, does not  change under this transformation. But the gauge condition for some other gauges, such as synchronous gauge, is not preserved in conformal transformation.
Considering the gauge in which conformal factor is uniform, i.e. $\delta \Omega=0$, all perturbation quantities become conformal invariant.
In single field model this condition is equivalent to comoving gauge, $\delta \Omega \propto \delta \phi$. However, in multiple field models it is different from comoving gauge.
We consider a multi-field model with scalar fields non-minimally coupled to gravity as follows:
\begin{equation}
S=\int {\rm d}^4x \, h \left[ \frac{1}{2} f(T, \phi^i)- \frac{1}{2} {\cal G}_{ij} \, \partial_\mu \phi^i \, \partial^\mu \phi^j -V(\phi^i) \right] , \label{G5}
\end{equation}
where $f(T,\phi^i)$ is an arbitrary function of torsion scalar and scalar fields, and $V(\phi^i)$ is a general potential.
So, we define the Lagrangian of scalar fields by
\begin{equation}
 {\cal L}_\phi:=- \frac{1}{2} {\cal G}_{ij} \, \partial_\mu \phi^i \, \partial^\mu \phi^j -V(\phi^i) .
\end{equation}
The presence of coupling between torsion and scalar fields in Eq.~\eqref{G5} implies that there is an energy transfer between  them. Consequently, we obtain
\begin{equation}
\nabla^\mu \Theta_{\mu \nu}^{(\phi)} = -\frac{1}{2} \frac{\partial f(T,\phi^i)}{\partial \phi^j} \, \nabla_\nu \phi^j , \label{b90}
\end{equation}
where
\begin{equation}
\Theta_{\mu \nu}^{(\phi)} =- \frac{2}{\sqrt{-g}} \frac{\delta \big( \sqrt{-g} {\cal L}_\phi \big)}{\delta g^{\mu \nu}}
\end{equation}
is the energy-momentum tensor of scalar fields.
Varying the action~\eqref{G5} with respect to the vierbein field leads to the following field equations:
\begin{align}
& -f_T G_{\mu \nu} + \frac{1}{2} ( f_T T -f ) g_{\mu \nu} +2 \partial_\alpha f_T S^{\phantom{\nu} \alpha}_{\nu \phantom{\alpha} \mu}
\nonumber \\	&- g_{\mu \nu} \left[ \frac{1}{2} {\cal G}_{ij} \partial_\alpha \phi^i \; \partial^\alpha \phi^j + V(\phi^i) \right]
\nonumber \\
&  +{\cal G}_{ij} \; \partial_\mu \phi^i \; \partial_\nu \phi^j =0. \label{field10}
\end{align}
The term $ 2 \partial_\alpha f_T S^{\phantom{\nu} \alpha}_{\nu \phantom{\alpha} \mu} $ in Eq.~\eqref{field10} leads to the violation of the local Lorentz symmetry. This term can, however, be removed by considering $ \partial_\alpha f_T=0 $, i.e. $f$ coincides with torsion scalar minimally coupled to scalar fields. This case is equivalent to General Relativity.
By introducing a new auxiliary field $\chi$, we can write the action~\eqref{G5} as
\begin{align}
S=\int {\rm d}^4x \, h \Bigg[ & \frac{1}{2} f_{,\chi}(\chi, \phi^i)\, T- \frac{1}{2} {\cal G}_{ij} \, \partial_\mu \phi^i \, \partial^\mu \phi^j
\nonumber \\	&-U(\chi,\phi^i) \Bigg] , \label{G7}
\end{align}
where the new potential is defined by $2U(\chi,\phi^i):=2V(\phi^i) -f(\chi,\phi^i) +f_\chi(\chi,\phi^i)\, \chi$. The variation of Eq. ~\eqref{G7} with respect to  the auxiliary field yields to
\begin{equation}
(T-\chi) f_{,\chi \chi}=0 .
\end{equation}
In the case $f_{,\chi \chi} \neq 0$, we get $\chi=T$.
In other words, $\chi$ comes from the nonlinearity of $f$ in terms of the torsion scalar.
Using a convenient conformal transformation in General Relativity, one can write scalar fields minimally coupled to the Ricci scalar. However, such transformation does not exist in teleparallel theories.
Performing the transformation~\eqref{G6}, torsion tensor transforms as follows
\begin{equation}
\hat{T}^\rho_{\phantom{\rho} \mu \nu}=T^\rho_{\phantom{\rho} \mu \nu}+ \Omega^{-1} \left( \delta^\rho_\nu \, \partial_\mu \Omega - \delta^\rho_\mu \, \partial_\nu \Omega \right) ,
\end{equation}
and torsion scalar becomes
\begin{equation}
T=\Omega^2 \hat{T} - 4 \, \Omega \, \hat{\partial}^\mu \Omega \, \hat{T}^\rho_{\phantom{\rho} \rho \mu} - 6 \, \hat{\partial}_\mu \Omega \, \hat{\partial}^\mu \Omega .
\end{equation}
We can also write transformation of the boundary term $B= 2 \partial_\mu \left( h T^{\rho \phantom{\rho} \mu}_{\phantom{\rho} \rho} \right) /h$ by
\begin{equation}
B=\Omega^2 \hat{B} -4 \Omega \hat{T}^{\rho \phantom{\rho} \mu}_{\phantom{\rho} \rho} \, \hat{\partial}_\mu \Omega -18 \, \hat{\partial}^\mu \Omega \, \hat{\partial}_\mu \Omega +6 \Omega \, \hat{\nabla}_\alpha \hat{ \nabla}^\alpha \Omega.
\end{equation}
Therefore, the action can be written as
\begin{align}
S=\int {\rm d}^4x \,  \hat{h} \bigg[& \frac{f_{,\chi}}{2 \Omega^2} \hat{T} - \frac{2f_{,\chi}}{ \Omega^3}  \hat{\partial}^\mu \Omega \, \hat{T}^\rho_{\phantom{\rho} \rho \mu}- \frac{3}{\Omega^4}  \hat{\partial}_\mu \Omega \, \hat{\partial}^\mu \Omega
\nonumber \\&-\frac{1}{2 \Omega^2} \, {\cal G}_{ij} \hat{\partial}_\mu \phi^i  \, \hat{\partial}^\mu \phi^j - \frac{ U(\chi, \phi^i) }{\Omega^4} \bigg]. \label{G8}
\end{align}
Plugging a conformal factor as $\Omega^2=f_{,\chi}$ and integrating, the action~\eqref{G8} gets the following form
\begin{align}
S=\int {\rm d}^4x \,  \hat{h} \bigg[& \frac{1}{2} \hat{T} + \frac{1}{2} \left( \ln f_{,\chi} \right) \hat{B}
-\frac{1}{2} \, \hat{{\cal G}}_{IJ} \hat{\partial}_\mu \phi^I  \, \hat{\partial}^\mu \phi^J
\nonumber \\	&- \hat{U}(\phi^I) \bigg], \label{action3}
\end{align}
where the scalar fields are $ \{\phi^I\}:= \{\phi^i, \chi\}  $, and $ \hat{U}(\phi^I):=U(\phi^I)/f_\chi^2 $ is the new potential.
We have also defined a new metric $ \hat{{\cal G}}_{IJ} $ of field space:
\begin{align}
\hat{{\cal G}}_{ij} =& \frac{1}{f_\chi} {\cal G}_{ij}+ \frac{3}{2} \frac{f_{,\chi i} f_{,\chi j}}{f_{,\chi}^2} ,
\nonumber \\
\hat{{\cal G}}_{i\chi} =& \hat{{\cal G}}_{\chi i} =  \frac{3}{2} \frac{f_{,\chi i} f_{,\chi \chi}}{f_{,\chi}^2} ,
\nonumber \\
\hat{{\cal G}}_{\chi \chi} =&  \frac{3}{2} \left( \frac{f_{,\chi \chi}}{f_{,\chi}} \right)^2 .
\end{align}
In the trivial case $f_{,\chi}=1$, the second term of Eq.~\eqref{action3} vanishes.
But in general we cannot find a conformal transformation to move to the Einstein frame.

Due to the conformal transformation, the matter Lagrangian becomes a function of scalar fields,
$ {\cal L}_{\rm m} \left[ \hat{h}^A_{\phantom{A} \mu} \Omega^{-1},\Psi_{\rm m} \right] $, where $\Psi_{\rm m}$ is all the matter fields.
This effect had been used in \emph{screening mechanisms}, e.g. chameleon \cite{chameleon} and symmetron \cite{symmetron}.
One can write
\begin{equation}
\hat{\Theta}^{({\rm m})}_{\mu \nu}= \Theta^{({\rm m})}_{\mu \nu} \Omega^{-2}.
\end{equation}
We also obtain
\begin{equation}
\hat{\nabla}_\mu \hat{\Theta}^{({\rm m})\mu}_{\, \, \nu}=\Omega^{-4} \left( \nabla_\mu \Theta^{({\rm m})\mu}_{\, \, \nu} - \Theta^{({\rm m}) \mu}_{\,\, \mu} \frac{\partial_\nu \Omega}{\Omega} \right).
\end{equation}
Considering the energy-momentum conservation for matter in the Jordan frame, i.e. $\nabla^\mu \Theta^{({\rm m})}_{\mu \nu}=0$\footnote{
This conservation law comes from the invariance of the matter action under the transformation of $ x^\mu \rightarrow x^\mu + \xi^\mu $.
}, in the new frame there will be an energy transfer between the torsion and scalar fields,
$ \hat{\nabla}^\mu \hat{\Theta}^{({\rm m})}_{\mu \nu} \neq 0 $.
However, for the traceless component of matter, this conservation is a conformal invariant.
In General Relativity the total energy-momentum tensor, i.e.
$ \hat{\Theta}^{({\rm m})}_{\mu \nu}+\hat{\Theta}^{(\phi)}_{\mu \nu} $, is conserved in the Einstein frame~\cite{Abedi}. However, it is simple to show that the total energy-momentum is not conserved for the action~\eqref{action3}.
The second term in the action~\eqref{action3} motivates us to use the boundary term in Eq.~\eqref{G5}.

Thus, we can modify the action through a function of scalar torsion, scalar fields and boundary term. In so doing, it is immediate to see that $ f(T,B) $ theories include both $f(R)$ and $f(T)$.
Indeed, if we consider an action of the form
\begin{equation}
S= \frac{1}{2}\int {\rm d}^4x \, h \left[  f(T,B, \phi^i)- {\cal G}_{ij} \, \partial_\mu \phi^i \, \partial^\mu \phi^j -2 V(\phi^i) \right] , \label{action4}
\end{equation}
with introducing four auxiliary fields $\chi_1$, $\chi_2$, $ \gamma_1 $ and $ \gamma_2 $,we can write:
\begin{align}
S=\frac{1}{2}\int {\rm d}^4x \, h  \Big[ & f(\chi_1, \chi_2, \phi^i) +  (T-\chi_1) \gamma_1
+ (B-\chi_2) \gamma_2
\nonumber \\	&- {\cal G}_{ij} \, \partial_\mu \phi^i \, \partial^\mu \phi^j
- 2 V(\phi^i) \Big] . \label{G10}
\end{align}
Its variation with respect to $\gamma_1$, $\gamma_2$, $\chi_1$ and $\chi_2$  yields to $ \chi_1=T $, $ \chi_2=B $, $ \gamma_1=f_{,\chi_1}(\chi_1, \chi_2) $ and $ \gamma_2=f_{,\chi_2}(\chi_1, \chi_2) $ respectively. Therefore, we  write Eq.~\eqref{G10} as
\begin{align}
S=\frac{1}{2}\int {\rm d}^4x \, h  \Big[ & f_{,\chi_1}  T+  f_{,\chi_2}  B-  {\cal G}_{ij} \, \partial_\mu \phi^i \, \partial^\mu \phi^j
\nonumber \\	&- 2U(\chi_1, \chi_2,\phi^i) \Big] ,
\end{align}
where $ 2U := 2 V  +f_{,\chi_1} \chi_1 + f_{,\chi_2}  \chi_2 -f $. We  now write this action by using the quantities defined in new frame:
\begin{align}
S=\int {\rm d}^4x \, \hat{h} \Bigg( & \frac{f_{,\chi_1}}{2 \Omega^2 }   \hat{T} + \frac{f_{,\chi_2}}{2 \Omega^2}  \hat{B}
-\frac{2 f_{,\chi_1}}{\Omega^3} \, \hat{\partial}_\mu \Omega \, \hat{T}^{\rho \phantom{\rho} \mu}_{ \phantom{\rho} \rho}
\nonumber \\ &	-\frac{3 f_{,\chi_1}}{\Omega^4}  \hat{\partial}_\mu \Omega \, \hat{\partial}^\mu \Omega -\frac{2 f_{,\chi_2}}{\Omega^3} \, \hat{\partial}_\mu \Omega \,
\hat{T}^{\rho  \phantom{\rho} \mu}_{ \phantom{\rho} \rho}
\nonumber \\
&
-\frac{9 f_{,\chi_2}}{\Omega^4}  \hat{\partial}_\mu \Omega \, \hat{\partial}^\mu \Omega
+\frac{3 f_{,\chi_2}}{\Omega^3} \hat{\nabla}_\alpha \hat{\nabla}^\alpha \Omega
\nonumber \\ &	-\frac{1}{2 \Omega^2} {\cal G}_{ij} \, \hat{\partial}_\mu \phi^i \, \hat{\partial}^\mu \phi^j
-\frac{U}{\Omega^4} \Bigg). \label{G9}
\end{align}
We can then collect terms that contain scalar fields coupled to the torsion tensor  as follows
\begin{equation}
- \int {\rm d}^4x \, \hat{h}\,\Omega^{-3} \left( 2 f_{, \chi_1} \, \hat{\partial}_\mu \Omega + \Omega \, \hat{\partial}_\mu f_{, \chi_2} \right) \, \hat{T}_{ \phantom{\rho}\rho}^{\rho  \phantom{\rho}  \mu}\ ,
\end{equation}
where we integrated by parts.

In order to omit  such a coupling, we can consider  $  f_{, \chi_1} \, \hat{\partial}_\mu \ln \Omega^2 = - \hat{\partial}_\mu f_{, \chi_2} $.
In analogy to the previous case, let us plug $ \Omega^2= f_{,\chi_1}$ in Eq.~\eqref{G9}, leading to a minimal coupling between the torsion scalar and the scalar fields, i.e.
\begin{align}
S=\int {\rm d}^4x \, \hat{h} \Bigg[& \frac{1}{2} \hat{T} + \frac{f_{,\chi_2}}{2 f_{,\chi_1}} \hat{B} - f_{,\chi_1}^{-2} \, \hat{\partial}_\mu ( f_{,\chi_1} +f_{,\chi_2} ) \, \hat{T}_{ \phantom{\rho} \rho}^{\rho  \phantom{\rho} \mu}
\nonumber \\ &-\frac{1}{2} \hat{{\cal G}}_{IJ} \, \hat{\partial}_\mu \phi^I \, \hat{\partial}^\mu \phi^J
+\hat{U}(\phi^I) \Bigg],
\end{align}
where the new  metric $\hat{{\cal G}}_{IJ}$ contains ${\cal G}_{ij}$ and all those terms that have $\hat{\partial }^\mu \Omega \; \hat{\partial }_\mu \Omega $ in the action~\eqref{G9}.
Considering $ f_{,\chi_2}= - f_{, \chi_1} $, scalar fields get completely decoupled from the torsion tensor.
The variation of Eq.~\eqref{action4} with respect to the vierbeins provides us the field equations
\begin{align}
& h^{ \phantom{A} \mu}_A\square f_B -h^{ \phantom{A} \nu }_A \nabla^\mu \nabla_\nu f_B +\frac{1}{2} B f_B h^{ \phantom{A} \mu }_A \nonumber \\&+2 \partial_\nu \left(  f_B+  f_T \right) S_A^{ \phantom{A} \nu \mu}
+ 2 h^{-1} \partial_\nu \left( h S_A^{ \phantom{A} \nu \mu} \right) f_T
\nonumber \\ &- 2 f_T T^\alpha_{ \phantom{\alpha} \nu A} S^{ \phantom{\alpha} \mu \nu} _\alpha -\frac{h}{2} h^{ \phantom{A} \mu}_A \Big[ f - {\cal G}_{ij} \partial_\alpha \phi^i \; \partial^\alpha \phi^j- 2 V(\phi^i)  \Big]=0,
\end{align}
or equivalently,
\begin{align}
& -f_T G_{\mu \nu} + \left( g_{\mu \nu} \square - \nabla_\mu \nabla_\nu \right) f_B \nonumber \\
& + \frac{1}{2} ( f_B B+ f_T T -f ) g_{\mu \nu} 	+2 S^{ \phantom{\nu} \alpha}_{\nu  \phantom{\alpha} \mu}\, \partial_\alpha (f_T+f_B)
\nonumber \\
& - g_{\mu \nu} \left[ \frac{1}{2} {\cal G}_{ij} \partial_\alpha \phi^i \; \partial^\alpha \phi^j + V(\phi^i) \right] 	\nonumber \\
&+{\cal G}_{ij} \; \partial_\mu \phi^i \; \partial_\nu \phi^j =0. \label{field9}
\end{align}
Thus, one can write again Eq.~\eqref{b90}.

The existence of the term $ 2 \partial_\alpha (f_T+f_B) S^{\; \alpha}_{\nu \; \mu}$ violates the local Lorentz symmetry.
The necessary  condition to obtain  the local Lorentz symmetry is again $f_T+f_B=c$, i.e. $f(T, B, \phi^i)= F(R,\phi^i)+c B$.
We note that the Ricci scalar constructed from the Weitzenb\"ock connection vanishes. In fact, we use $R$ to refer to the Ricci scalar constructed from the Levi-Civita connection.
In this case $f_T=-F_R$ and $f_B=F_R+c$, consequently Eq.~\eqref{field9} reduces to the field equation proper of a $F(R)$-theory,
\begin{align}
& F_R G_{\mu \nu} + \left( g_{\mu \nu}\square - \nabla_\mu \nabla_\nu \right) F_R + \frac{1}{2} (F_R R-F ) g_{\mu \nu} 	\nonumber \\
&	- g_{\mu \nu} \left[ \frac{1}{2} {\cal G}_{ij} \partial_\alpha \phi^i \; \partial^\alpha \phi^j + V(\phi^i) \right] \nonumber \\
&+{\cal G}_{ij} \; \partial_\mu \phi^i \; \partial_\nu \phi^j =0\ .
\end{align}
It is worth noticing that the gravitational coupling in $F(R,\phi^i)$ has  been extensively investigated. Thus, we do not need to furnish any further details on it. The torsion scalar and the boundary term depend on first and second derivatives of the vierbein fields respectively. The second term in Eq.~\eqref{field9} contains  $ f_{BB} \, \partial_\mu \partial_\nu B $. Hence, the field equations turn out to be fourth order. In the case of a Lagrangian of the form
\begin{equation}
f(T,B,\phi) = A(T,\phi)\, B + C(T,\phi),
\end{equation}
this term vanishes. However, the fourth term contains the third derivatives of the vierbein fields. Considering $A(T,\phi)=c$ where $c$ is a constant, one gets a second-order theory with the Lagrangian
\begin{equation}
f(T,B,\phi)=c B + C(T,\phi).
\end{equation}
\noindent Thus, if the Lagrangian is a function of the torsion scalar non-minimally coupled to scalar field, the whole scenario becomes a second-order theory.


\section{Scalar field  non-minimally coupled to torsion scalar and the boundary term}
\label{STB}

In this section, we obtain the effective gravitational coupling with a scalar field  non-minimally coupled to both torsion scalar and the boundary term. We restrict our attention to the following action, introduced in~\cite{Wright1}:
\begin{align}
S=\int {\rm d}^4x \, h \Bigg[&-\frac{T}{2} -\frac{1}{2} \Big( \partial_\mu \phi \; \partial^\mu \phi +\xi T \phi^2
\nonumber \\&
+\chi B \phi^2
\Big)-V(\phi) +{\cal L}_{\rm m} \Bigg] . \label{action5}
\end{align}
The case in which $\xi=\chi=0$ reduces to General Relativity. If $\xi+\chi=0$, the scalar field is non-minimally coupled to the Ricci scalar.

The variation of the action with respect to vierbein fields and scalar field yields to the following field equations respectively:
\begin{align}
&( 1+\xi \phi^2) G^\mu_\nu - \delta^\mu_\nu \left[ \frac{1}{2} \partial_\alpha \phi  \; \partial^\alpha \phi-V(\phi)\right] +\nabla^\mu \phi \, \nabla_\nu \phi
\nonumber \\
&
-4(\xi+\chi) S_\nu^{ \phantom{\nu} \lambda \mu} \phi \, \partial_\lambda \phi
- \chi \left( \delta^\mu_\nu \square - \nabla^\mu \nabla_\nu  \right)\phi^2 \nonumber \\ &=\Theta^\mu_\nu\ , \label{b11}
\end{align}
\begin{equation}
\nabla_\mu \nabla^\mu \phi + V_\phi = (\xi T+ \chi B) \phi\ . \label{b12}
\end{equation}
In particular, Eq. \eqref{b12} in the background spacetime takes the form
\begin{equation}
\ddot{\phi} +3H\dot{\phi}+6 \Big[ \xi H^2+ \chi(3H^2+\dot{H)} \Big] \phi +V_\phi=0 .
\end{equation}
Further, Einstein's field equations can be written as follows
\begin{align}
3H^2 =& \rho_{\rm m}+\rho_\phi , \nonumber \\
3H^2+2\dot{H} =&  -(p_{\rm m}+p_\phi) ,
\end{align}
where
\begin{align}
\rho_\phi =& \frac{\dot{\phi}^2}{2} +V(\phi) -3\xi H^2 \phi^2 +6 \chi H \phi \dot{\phi} , \nonumber \\
p_\phi =& \frac{1}{2} (1-4\chi) \dot{\phi}^2 - V(\phi) +2 H\phi \dot{\phi} (2\xi +3\chi) \nonumber \\ &+3H^2 \phi^2 (\xi+8\chi^2)+2\phi^2 \dot{H} (\xi+6\chi^2)
\nonumber \\ &
+2\chi \phi V_\phi(\phi)  .
\end{align}
Up to  first-order perturbations, the boundary term becomes
\begin{align}
B=&-18H^2-6\dot{H} +12(3H^2+ \dot{H}) \psi +6H (6\dot{\varphi}+\dot{\psi})
\nonumber \\ &
+6\ddot{\varphi}
+2 \frac{\partial^2}{a^2} (2\zeta+\psi-2\varphi).
\end{align}
Hereafter we use the quasi-static approximation, \textit{i.e.} $ |\dot{Q} | \leq |HQ| $ for $Q= \phi, \psi, \varphi$ on sub-horizon scales $ k\gg aH$.  Hence, we get
\begin{align}
B\simeq& -6(3H^2+\dot{H})+2 a^{-2} \partial^2 (2\zeta+\psi-2\varphi), \nonumber \\
T\simeq& -6H^2+4 a^{-2}  \partial^2 \zeta,
\nonumber \\
R \simeq&-6(2H^2+\dot{H})+ 2 a^{-2} \partial^2  (\psi-2\varphi).
\end{align}
The antisymmetric part of the field equation is
\begin{equation}
4 (\xi + \chi) \Big( g^{\mu \alpha} S^{ \phantom{\mu} \lambda \beta}_\mu - g^{\nu \beta} S^{ \phantom{\nu} \lambda \alpha}_\nu  \Big) \phi \; \partial_\lambda \phi =0.  \label{E1}
\end{equation}
This constraint  in FLRW background  is automatically satisfied. So, we consider below two different cases:
\subsection{$\xi+\chi=0$}

The first case lies on assuming Eq.~\eqref{E1} satisfied at all perturbed orders for $\xi+\chi=0$. In this case, as the scalar field is non-minimally coupled to the Ricci scalar, the field equations are locally-Lorentz invariant. This model has extensively been studied in~\cite{Boisseau}.
Obtaining the equation for the matter perturbations growth implies that one gets the effective gravitational coupling as:
\begin{equation}
G_{\rm eff} \simeq  \frac{1}{1+\xi \phi^2} . \label{Gef1}
\end{equation}

\subsection{$\xi+\chi \neq 0$}

The second possibility is $ \chi \neq -\xi $.
The constraint~\eqref{E1} at first order of perturbations leads to
\begin{equation}
\varphi= - \frac{H}{\dot{\phi}} \delta \phi \label{A1} .
\end{equation}
whereas the $00$-component of field equations is
\begin{widetext}
	\begin{align}
	&-6H^2 \xi \phi \, \delta \phi +(1+ \xi \phi^2) \bigg[ 6H(\dot{\varphi}+H\psi)+ 2 \frac{\partial^2}{a^2} \varphi \bigg] 	-  \psi \dot{\varphi}^2
	+\dot{\phi} \, \dot{\delta \phi} + V_\phi \, \delta \phi - \chi \Bigg[ 2(\ddot{\phi} \, \delta \phi+ 2 \dot{\phi} \, \dot{\delta \phi} + \phi \, \delta \ddot{\phi} 	\nonumber \\ & - \dot{\psi} \phi \ddot{\phi})
	-4\psi (\dot{\phi}^2+\phi \ddot{\phi})
	-\frac{2}{a^2} \Big\{ \phi\partial^2 \, \delta \phi -3a^2 H (\dot{\phi}\, \delta \phi + \phi \, \dot{ \delta \phi})
 +3a^2 \phi \dot{\phi} [2H(\psi+\varphi)+ \dot{\varphi}] \Big\} +12\phi \dot{\phi} H \varphi -2 \ddot{\phi} \, \delta \phi 	\nonumber \\ &-4 \dot{\phi} \, \dot{\delta \phi}
	-2 \phi \, \delta \ddot{\phi} + 2 \dot{\psi} \phi \dot{\phi} +4 (\dot{\phi}^2 + \phi \ddot{\phi}) \psi  \Bigg]
	=- \delta \rho_{\rm m} \ .
	\end{align}
\end{widetext}
Deeply inside the Hubble radius, and for a slow-varying potential, the former equation becomes
\begin{equation}
2(1+\xi \phi^2) \left( \frac{k}{a} \right)^2 \varphi  +2 \chi \phi \left( \frac{k}{a} \right)^2 \delta \phi =  \delta \rho_{\rm m}.  \label{A2}
\end{equation}
We can also write the $ij$-component of Eq.~\eqref{b11} with $i \neq j$ as
\begin{align}
&(1+\xi \phi^2)  \frac{ \partial^i \partial_j}{a^2} (\psi - \varphi) +2 (\xi +\chi)  \phi \dot{\phi}   \frac{ \partial^i \partial_j}{a} \alpha \nonumber \\ &-2 \chi \phi  \frac{ \partial^i \partial_j}{a^2} \delta \phi =0.
\end{align}
This equation can be written by:
\begin{equation}
(1+\xi \phi^2) (\psi - \varphi) +2 (\xi +\chi)  \phi \dot{\phi} a  \alpha -2 \chi \phi  \delta \phi =0 \label{A4}.
\end{equation}
We can also write $\delta \Theta_0^{\,i}$  and Eq.~\eqref{b12} at first order of perturbations,
\begin{widetext}
	\begin{align}
	& \frac{2}{a^2} (1+\xi \phi^2) \, \partial^i (\dot{\varphi}+ H\psi) -\frac{1}{a^2} \dot{\phi} \,\partial^i  \delta \phi
+ \frac{4}{a^2} (\xi+\chi) \phi \dot{\phi} \left(\partial^i \varphi + \frac{1}{2} \partial_l B^{li}\right)
	+ \frac{ 2 \chi}{a^2} \big( -\dot{\phi} \, \partial^i  \delta\phi - \phi \, \partial^i  \dot{\delta \phi} - \phi \dot{\phi} \, \partial^i \psi
	+ H\phi \, \partial^i \delta \phi \big)  \nonumber \\
	&= -\frac{1}{a^2} (\rho_{\rm m}+p_{\rm m}) \partial^i \delta u .
	\\
	& \ddot{\delta \phi} + 3 H\dot{\delta \phi} + \left( V_{\phi \phi} - \frac{\partial^2}{a^2} \right) \delta \phi + 6 \xi H^2 \delta \phi -2 (\ddot{\phi} + 3H \dot{\phi} ) \psi
		- \dot{\phi} (3 \dot{\varphi} + \dot{\psi} ) - 12 \xi H \phi (\dot{\varphi} + H\psi ) -4 \xi \phi \frac{\partial^2}{a^2} \zeta
	\nonumber \\
	& +6\chi \, \delta \phi (3H^2+ \dot{H}) -\chi \phi \Big[ -18H^2-6\dot{H}
		+12(3H^2+ \dot{H}) \psi +6H (6\dot{\varphi}+\dot{\psi})+6\ddot{\varphi}
	+2 \frac{\partial^2}{a^2} (2\zeta+\psi-2\varphi) \Big]=0 . \label{b13}
	\end{align}
\end{widetext}
Then, Eq.~\eqref{b13} becomes
\begin{equation}
\left( \frac{k}{a} \right)^2  \delta \phi + 4 \xi \phi \left( \frac{k}{a} \right)^2 \zeta + 2 \chi \phi \left( \frac{k}{a} \right)^2 (2\zeta +\psi -2\varphi) =0 \label{A3}.
\end{equation}
Using Eqs.~\eqref{A1}, \eqref{A2}, \eqref{A4} and \eqref{A3},  we can write  the matter density perturbations in which the effective gravitational coupling takes the form
\begin{align}
G_{\rm eff}\simeq \left( 1+\xi \phi^2 -4\frac{\chi \phi \dot{\phi}}{H}-\frac{\dot{\phi}^2}{2H^2} \right)  \left( 1+\xi \phi^2 - \frac{\chi \phi \dot{\phi}}{H} \right)^{-2} .  \label{Gef2}
\end{align}
For pure teleparallel theory, provided by $\chi=0$, it becomes
\begin{align}
G_{ \rm eff}\simeq
\frac{1}{1+\xi \phi^2} \left( 1-\frac{\dot{\phi}^2}{2H^2(1+\xi \phi^2)} \right) .
\label{G_eff}
\end{align}
The above result confirms the outcomes discussed in \cite{Geng3}. Finally, we can also obtain the effective gravitational coupling when $ \xi=0 $, which reads
\begin{align}
G_{\rm eff}\simeq \left( 1-4\frac{\chi \phi \dot{\phi}}{H}-\frac{\dot{\phi}^2}{2H^2} \right)  \left( 1 - \frac{\chi \phi \dot{\phi}}{H} \right)^{-2} .
\end{align}


\section{Comparison with observations}
\label{observations}

\begin{figure}[h!]
\begin{center}
\includegraphics[width=3.3in]{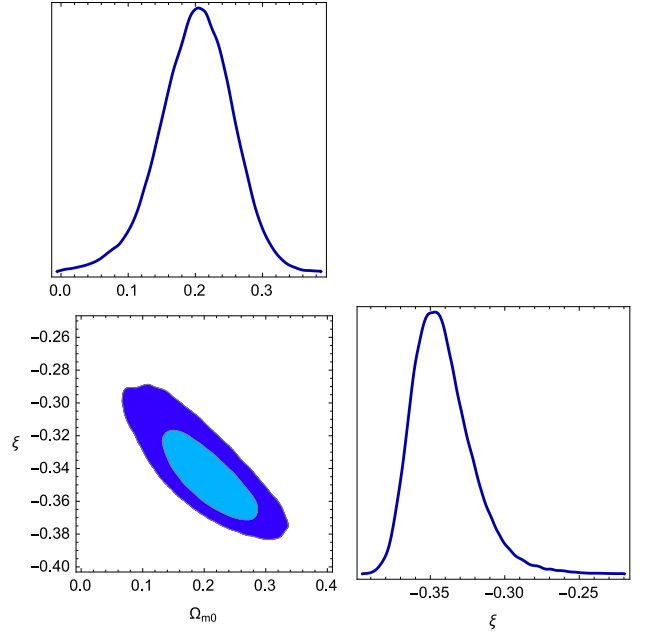}
\caption{68\% and 95\% confidence level contours and posterior distributions as result from the MCMC analysis on the Hubble rate data.}
\label{fig:xi}
\end{center}
\end{figure}

In this section, we compare the growth of matter perturbations for a specific form of teleparallel theory with the predictions of the concordance $\Lambda$CDM model. Furthermore, we confront the theoretical predictions with the latest available data.
In particular, we consider action \eqref{action1} with
\begin{equation}
f(T,\phi,X)=T + \partial_\mu \phi \ \partial^\mu \phi +\xi T \phi^2.
\label{f(T,phi)}
\end{equation}
Observational tests on this model have been performed in \cite{Gu12}, where a best-fit value of $\xi=-0.35$ was found.
We can study the dynamics of such a model by introducing the variables $x\equiv \dot{\phi}/\sqrt{6}H$ and $y \equiv \sqrt{-\xi}\phi$ satisfying the Friedmann equation $\Omega_{\rm m}+x^2+y^2=1$ \cite{Geng3}. The evolution of these variable with respect to the scale factor is given by
\begin{align}
&a\dfrac{{\rm d}x}{{\rm d}a}=-\left(3+\dfrac{\dot{H}}{H^2}\right)x+\sqrt{-6\xi}\ y \\
&a\dfrac{{\rm d}y}{{\rm d}a}=\sqrt{-6\xi}\ x
\end{align}
where
\begin{equation}
\dfrac{\dot{H}}{H^2}=\dfrac{1}{1-y^2}\left[-3x^2+2\sqrt{-6\xi}\ x y -\dfrac{3}{2}\Omega_{\rm m}\right].
\end{equation}
The above system can be solved numerically by using $\Omega_m=0.999$ and $y= 10^{-6}$ as initial conditions for $a\ll 1$.
We can write the evolution of the Hubble rate as function of the redshift in the form
\begin{equation}
\dfrac{H}{H_0}=\sqrt{\Omega_{\rm m0}(1+z)^{3}+\Omega_{\rm DE,0}(1+z)^{3\left(1+\omega_{\rm DE}\right)}}\ ,
\label{eq:H(a)}
\end{equation}
where $\Omega_{\rm DE,0}=1-\Omega_{\rm m0}$. We emphasize that $\omega_{\rm DE}$ is a time-dependent function that evolves according to
\begin{equation}
\omega_{\rm DE}=\dfrac{x\left(-3x+4y\sqrt{-6\xi}\right)}{3\left(-1+y^2\right)\left(x^2+y^2\right)}\ .
\end{equation}
Eq.~\eqref{eq:H(a)} can be tested against data to find constraints on the coupling constant $\xi$. In particular, we here consider the most recent model-independent measurements of the Hubble rate obtained through the differential age method (see \cite{Jimenez02} for details).  We list the data together with the corresponding references in Table~\ref{tab:OHD} in the Appendix. The Likelihood function reads
\begin{equation}
\mathcal{L}\propto\exp\left\{-\dfrac{1}{2}\sum_{i=1}^{31}\left(\dfrac{H_{th}(z_i)-H_{obs}(z_i)}{\sigma_{H,i}}\right)^2\right\}.
\end{equation}
We, thus, performed a Markov Chain Monte Carlo (MCMC) integration via the Metropolis algorithm for parameters estimation. Fixing $H_0=70$ km/s/Mpc, we assumed uniform priors for the fitting parameters: $\Omega_{\rm m0}\in(0,1)$ and $\xi\in(-1,0)$. Our numerical analysis provides $\Omega_{\rm m0}=0.218 \pm 0.054$ and $\xi=-0.351 \pm 0.020$, which is consistent with the result obtained in \cite{Gu12}. In Fig.~\ref{fig:xi}, we show the $1\sigma$ and $2\sigma$ confidence regions with the 1D posterior distributions.

We want now to study the evolution of the perturbations on sub-horizon scales. To do that,  we introduce the growth rate $g\equiv {\rm d}\ln \delta_m/{\rm d}\ln a$, so that the matter density perturbation equation can be written as
\begin{equation}
a \dfrac{{\rm d}g}{{\rm d}a}+\left(2+\dfrac{\dot{H}}{H^2}\right)g+g^2=\dfrac{3}{2}G_\text{eff}\ \Omega_{\rm m}\ ,
\label{g(a)}
\end{equation}
where (cf. Eq.~\eqref{G_eff})
\begin{equation}
G_\text{eff}=\dfrac{1}{1-y^2}\left[1-\dfrac{6 x^2}{2(1-y^2)}\right].
\end{equation}
Eq.~\eqref{g(a)} can be solved numerically by using $g=1$  as initial condition for $a\ll 1$. Once $g(a)$ is obtained, we can find the matter density contrast by integrating
\begin{equation}
\dfrac{a}{\delta_{\rm m}}\dfrac{{\rm d}\delta_{\rm m}}{{\rm d}a}=g(a)
\end{equation}
with the initial condition $\delta_{\rm m}(a\ll 1)=a$.

We can estimate the deviations from the General Relativity case by comparing our results with the $\Lambda$CDM scenario. The perturbation equation for the matter density contrast in the $\Lambda$CDM model is given by
\begin{equation}
\dfrac{{\rm d}^2\delta_{\rm m}}{{\rm d}a^2}+\left[\dfrac{3}{a}+\dfrac{{\rm d}E/{\rm d}a}{E}\right]\dfrac{{\rm d}\delta_{\rm m}}{{\rm d}a}=\dfrac{3}{2}\dfrac{\Omega_{{\rm m}0}}{a^5 E^2}\delta_{\rm m}\ ,
\end{equation}
where
\begin{equation}
E(a)=\sqrt{\Omega_{{\rm m}0}a^{-3}+\Omega_\Lambda}
\end{equation}
with the constraint $\Omega_{{\rm m}0}+\Omega_\Lambda=1$.

\begin{figure}[h!]
\begin{center}
\includegraphics[width=3.3in]{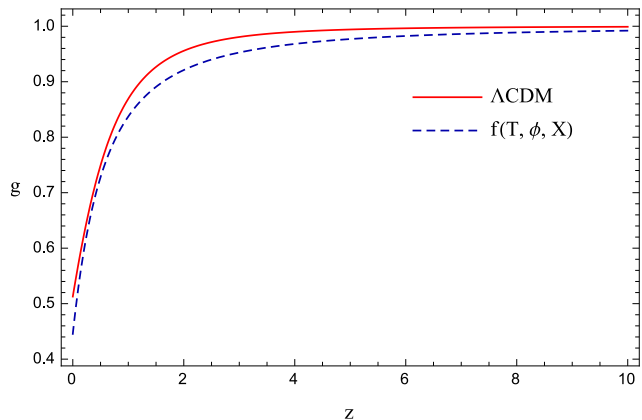}
\caption{Growth rate as function of the redshift for the model of Eq.~\eqref{f(T,phi)} with $\xi=-0.351$ (blue dashed line) and for the $\Lambda$CDM model with $\Omega_{{\rm m}0}=0.3$ (red solid line).}
\label{fig:g}
\end{center}
\end{figure}

\begin{figure}[h!]
\begin{center}
\includegraphics[width=3.3in]{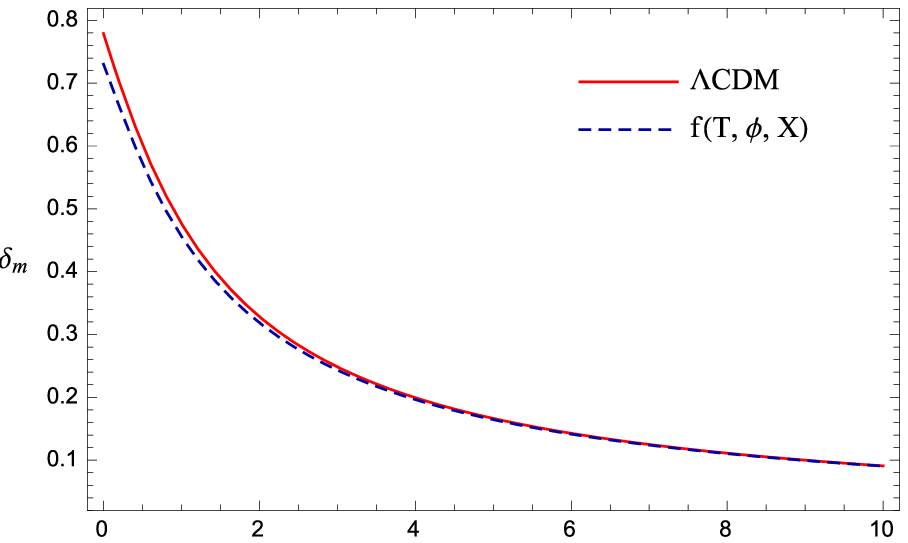}\\
\vspace{0.3cm}
\includegraphics[width=3.3in]{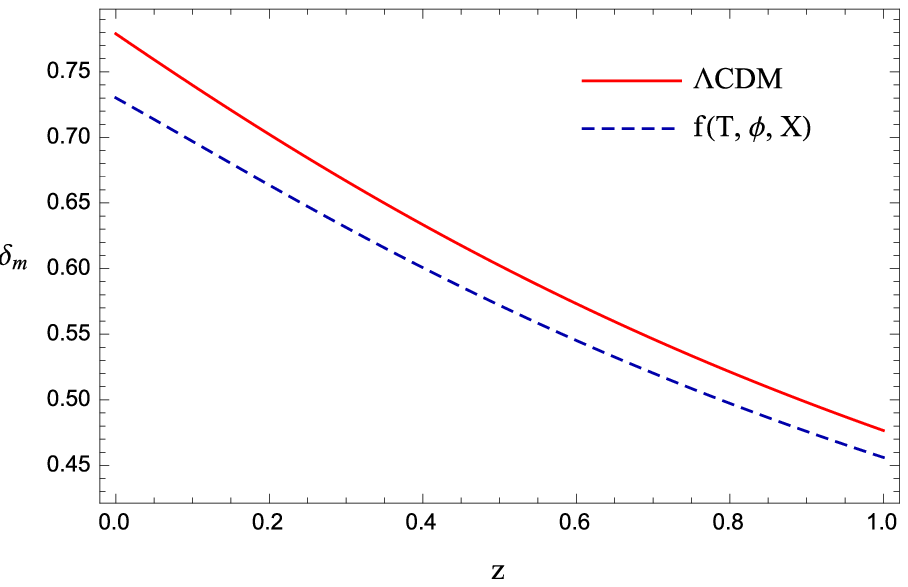}
\caption{Evolution of the matter density contrast as a function of the redshift for the model of Eq.~\eqref{f(T,phi)} with $\xi=-0.351$ (blue dashed line) and for the $\Lambda$CDM model  with $\Omega_{{\rm m}0}=0.3$ (red solid line). The bottom panel zooms in on the interval $z\in[0,1]$.}
\label{fig:delta}
\end{center}
\end{figure}

Using the best-fit results of our MCMC analysis for the $f(T,\phi,X)$ model, we show in Fig.~\ref{fig:g} the behaviour of the growth rate resulting from the solution of Eq.~\eqref{g(a)} compared to the $\Lambda$CDM model with $\Omega_{{\rm m}0}=0.3$. We observe a slower growth rate for the $f(T,\phi,X)$ model with respect to $\Lambda$CDM, which means that the gravitational interaction is weaker than in the General Relativity case. Our result is in agreement with previous findings obtained in \cite{Dent,Fu11}.
In Fig.~\ref{fig:delta}, we also show the evolution of the matter density contrast for the two models.

Over recent years, measurements from redshift space distortion and weak lensing have been obtained in the redshift interval $0.02\leq z \leq 1.4$ for the factor
\begin{equation}
g\sigma_8(z)\equiv g(z)\sigma_8(z)\ ,
\end{equation}
where $\sigma_8(a)=\sigma_8\frac{\delta(a)}{\delta(1)}$ is the r.m.s. fluctuation of the linear density field inside a radius of $8h^{-1}$Mpc, and $\sigma_8$ is its present day value.
A collection of these measurements is presented in \cite{Nesseris17}. The data points we consider here are summarized in Table~\ref{tab:GRF data} in the Appendix. In Fig.~\ref{fig:gsigma8}, we show the functional behaviour of $g\sigma_8(z)$ for the model of Eq.~\eqref{f(T,phi)} and for the $\Lambda$CDM model with the indicative values of $\Omega_{{\rm m}0}=0.3$ and $\sigma_8=0.8$.

\begin{figure}[h!]
\begin{center}
\includegraphics[width=3.3in]{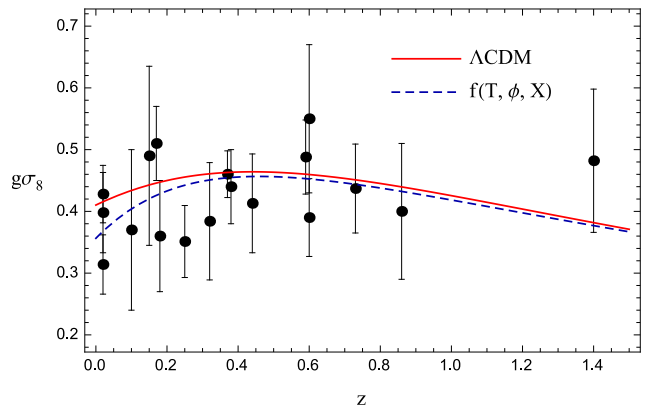}
\caption{Growth rate factor as function of the redshift for the model of Eq.~\eqref{f(T,phi)} with $\xi=-0.351$ (blue dashed line) and for the $\Lambda$CDM model with $\Omega_{{\rm m}0}=0.3$ (red solid line). We fix $\sigma_8=0.8$.}
\label{fig:gsigma8}
\end{center}
\end{figure}

\newpage
\section{Conclusion and perspectives}
\label{conclusion}

In this work, we have studied matter density perturbations in a  generalized theory of teleparallel gravity, with Lagrangian $f(T,\phi,X)$. In particular, we investigated a Lagrangian depending upon a scalar field $\phi$, its kinetic term $X$ and on the torsion scalar $T$. We derived the corresponding background field equations and then investigated scalar and vector perturbations. We found that the extra degrees of freedom have no significant effect on small scales. In analogy to $f(R)$ and $ f(T) $-theories, the effective gravitational coupling is only modified for scalar perturbations, and vector degrees of freedom remain decaying modes. We thus derived the gravitational coupling in multiple scalar field model. The Lagrangian considered in the present work covers many teleparallel models. However, our approach can be extended to more general frameworks that include Gauss-Bonnet term \cite{ Kofinas, Kofinas1, Bahamonde5}, Galileon \cite{galileon1,galileon2}, Proca theories \cite{proca} and other modified Lagrangian. Conformal transformations have been studied for $f(T)$ theory non-minimally coupled to scalar fields. In the general form, we showed that it has no Einstein frame associated to it. However, one can obtain the Einstein frame and local Lorentz symmetry in existence of the boundary term $B$ when its Lagrangian coincides with $f(R)$. The action~\eqref{action5} is equivalent to consider the scalar field coupled to gravity by
$(\chi R+\sigma T)\phi^2$  or $ (-\xi R + \sigma B) \phi^2 $ where $\sigma :=\xi+\chi$. As $\sigma =0$ we recover the local Lorentz symmetry and the effective gravitational coupling is given by Eq.~\eqref{Gef1}. For $\sigma \neq 0$, we obtained the coupling got in Eq.~\eqref{Gef2}. To check the goodness of our theoretical landscape, we computed the growth rate of matter perturbations for the specific model $f(T,\phi,X)=T + \partial_\mu \phi\  \partial^\mu \phi +\xi T \phi^2$. Fixing the coupling to a constant value $\xi$ by using the best-fit value suggested by cosmological measurements, we  compared our outcomes with the most recent bounds and with the $\Lambda$CDM predictions. We found that the growth rate turns out to be slower for $f(T,\phi,X)$ theories than the one predicted by the concordance model. We noticed that this is due to a weaker effective gravitational interaction compared with Newtonian's gravity. Our result is consistent with previous outcomes recently obtained in the literature. Future works will be dedicated to work on the evolution of perturbations on super-horizon scales. There, one expects to find much more significant deviations from General Relativity. To this end, we will even work on Monte Carlo simulated analyses combining low and high redshift data surveys, with the aim to get more stringent constraints on the coupling $\xi$ predicted by our  $f(T,\phi,X)$ approach.

\section*{Acknowledgements}
This paper is based upon work from COST action CA15117 (CANTATA), supported by COST (European Cooperation in Science and Technology).


\clearpage

\appendix*
 \section{Experimental data compilations}
 \label{sec:appendix}

In this appendix, we list the observational Hubble data and the growth rate measurements used in this work.
\begin{table}[h]
\small
\begin{center}
\setlength{\tabcolsep}{1.5em}
\begin{tabular}{c c c }
\hline
\hline
 $z$ &$H \pm \sigma_H$ &  References \\
\hline
0.0708	& $69.00 \pm 19.68$ & \cite{Zhang14} \\
0.09	& $69.0 \pm 12.0$ & \cite{Jimenez02} \\
0.12	& $68.6 \pm 26.2$ & \cite{Zhang14} \\
0.17	& $83.0 \pm 8.0$ & \cite{Simon05} \\
0.179 & $75.0 \pm	4.0$ & \cite{Moresco12} \\
0.199 & $75.0	\pm 5.0$ & \cite{Moresco12} \\
0.20 &$72.9 \pm 29.6$ & \cite{Zhang14} \\
0.27	& $77.0 \pm 14.0$ & \cite{Simon05} \\
0.28	& $88.8 \pm 36.6$ & \cite{Zhang14} \\
0.35	& $82.1 \pm 4.85$ & \cite{Chuang12}\\
0.352 & $83.0	\pm 14.0$ & \cite{Moresco16} \\
0.3802	& $83.0 \pm 13.5$ & \cite{Moresco16}\\
0.4 & $95.0	\pm 17.0$ & \cite{Simon05} \\
0.4004	& $77.0 \pm 10.2$ & \cite{Moresco16} \\
0.4247	& $87.1 \pm 11.2$  & \cite{Moresco16} \\
0.4497 &	$92.8 \pm 12.9$ & \cite{Moresco16}\\
0.4783	 & $80.9 \pm 9.0$ & \cite{Moresco16} \\
0.48	& $97.0 \pm 62.0$ & \cite{Stern10} \\
0.593 & $104.0 \pm 13.0$ & \cite{Moresco12} \\
0.68	& $92.0 \pm 8.0$ & \cite{Moresco12} \\
0.781 & $105.0 \pm 12.0$ & \cite{Moresco12} \\
0.875 & $125.0 \pm 17.0 $ & \cite{Moresco12} \\
0.88	& $90.0 \pm 40.0$ & \cite{Stern10} \\
0.9 & $117.0 \pm 23.0$ & \cite{Simon05} \\
1.037 & $154.0 \pm 20.0$ & \cite{Moresco12} \\
1.3 & $168.0 \pm 17.0$ & \cite{Simon05} \\
1.363 & $160.0 \pm 33.6$ & \cite{Moresco15} \\
1.43	& $177.0 \pm18.0$ & \cite{Simon05} \\
1.53	& $140.0	\pm 14.0$ & \cite{Simon05} \\
1.75	 & $202.0 \pm 40.0$ & \cite{Simon05} \\
1.965& $186.5 \pm 50.4$ & \cite{Moresco15} \\
\hline
\hline
\end{tabular}
\caption{Differential age Hubble data used in this work. The Hubble rate is given in units of km/s/Mpc.}
 \label{tab:OHD}
\end{center}
\end{table}

\begin{table}[h!]
\small
\begin{center}
\setlength{\tabcolsep}{1.5em}
\begin{tabular}{c c c }
\hline
$z$ & $g\sigma_8(z)$ & References \\
\hline
0.02 & $0.428 \pm 0.0465$  & \cite{Huterer17}\\
0.02 & $0.389 \pm 0.065$ & \cite{Turnbull12, Hudson12}\\
0.02 & $0.314 \pm 0.048$ &  \cite{Davis11, Hudson12} \\
0.10 & $0.370 \pm 0.130$ &  \cite{Feix15}\\
0.15 & $0.490 \pm 0.145$ &  \cite{Howlett15}\\
0.17 & $0.510 \pm 0.060$ &  \cite{Song09}\\
0.18 & $0.360 \pm 0.090$ &  \cite{Blake13}\\
0.38 & $0.440 \pm 0.060$ &  \cite{Blake13}\\
0.25 & $0.3512 \pm 0.0583$  & \cite{Samushia12}\\
0.37 & $0.4602 \pm 0.0378$  & \cite{Samushia12}\\
0.32 & $0.384 \pm 0.095$ &  \cite{Sanchez14}\\
0.59 & $0.488 \pm 0.060$ &  \cite{Chuang16}\\
0.44 & $0.413 \pm 0.080$ &  \cite{Blake12}\\
0.60 & $0.390 \pm 0.063$  & \cite{Blake12}\\
0.73 & $0.437 \pm 0.072$ &  \cite{Blake12}\\
0.60 & $0.550 \pm 0.120$ &  \cite{Pezzotta16}\\
0.86 & $0.400 \pm 0.110$ & \cite{Pezzotta16}\\
1.40 & $0.482 \pm 0.116$ & \cite{Okumura16}\\
\hline
\end{tabular}
\caption{Growth rate measurements considered in this work.}
\label{tab:GRF data}
\end{center}
\end{table}


\clearpage

\begin{thebibliography}{99}

\bibitem{Perlmutter} S. Perlmutter et. al.,   \href{http://dx.doi.org/10.1086/307221}{Astrophys. J. \textbf{517}, 565 (1999)}.

\bibitem{Tegmark} M. Tegmark et. al.,   \href{http://dx.doi.org/10.1103/PhysRevD.69.103501}{ Phys. Rev.  D \textbf{69}, 103501  (2004)}.

\bibitem{Allen} S. W.  Allen  et. al.,  \href{http://dx.doi.org/10.1111/j.1365-2966.2004.08080.x}{Mon. Not. R. Astron. Soc.   \textbf{353}, 457 (2004)}.

\bibitem{Planck} Planck Collaboration (P.A.R. Ade et. al.) \href{http://dx.doi.org/10.1051/0004-6361/201321580}{Astron. Astrophys. \textbf{571},  A12 (2014)}.

\bibitem{orl} A. Aviles, C. Gruber, O. Luongo, H. Quevedo,      \href{http://dx.doi.org/10.1103/PhysRevD.86.123516}{Phys. Rev. D {\bf 86}, 123516 (2012)}.



\bibitem{LiWang} M. Li, X. D. Li, S. Wang and Y. Wang,  \href{http://dx.doi.org/10.1007/s11467-013-0300-5}{ Front. Phys. (Beijing)  {\bf 8}, 828-846 (2013)}.

\bibitem{Hayashi} K. Hayashi  and T. Shirafuji,  \href{http://dx.doi.org/10.1103/PhysRevD.19.3524}{Phys. Rev.  D \textbf{19}, 3524 (1979)}.

\bibitem{Hayashi2} K. Hayashi  and T. Shirafuji,    \href{https://doi.org/10.1103/PhysRevD.24.3312}{Phys. Rev. D \textbf{24},  3312 (1981)}.

\bibitem{Aldrovandi} R. Aldrovandi and J. G. Pereira, Teleparallel Gravity, An Introduction (Springer, 2013).

\bibitem{Hehl}  F. W. Hehl, J. D. McCrea, E. W. Mielke and Y. Ne\'eman,  \href{http://dx.doi.org/10.1007/BF01883159}{Found. Phys.  \textbf{19}, 1075 (1989)}.

\bibitem{Maluf}  J. W. Maluf,   \href{http://dx.doi.org/10.1002/andp.201200272}{Ann. Phys.  (Berlin)  \textbf{525}, 339 (2013)}.

\bibitem{Abedi3} H. Abedi, M. Wright and A. M. Abbassi,  \href{http://dx.doi.org/10.1103/PhysRevD.95.064020}{Phys. Rev. D {\bf 95}, 064020  (2017)}.


\bibitem{orl1}
A. Aviles, A. Bravetti, S. Capozziello, O. Luongo, \href{http://dx.doi.org/10.1103/PhysRevD.87.064025}{Phys. Rev. D {\bf 87}, 064025 (2013)}.

\bibitem{orl2}
S. Capozziello, O. Luongo, E. N. Saridakis, \href{http://dx.doi.org/10.1103/PhysRevD.91.124037}{Phys. Rev. D {\bf 91}, 124037 (2015)}.

\bibitem{rocco}
S. Capozziello, R. D'Agostino, O. Luongo,  \href{http://dx.doi.org/10.1007/s10714-017-2304-x}{Gen. Rel. Grav. \textbf{49}, 141 (2017)}.


\bibitem{Bengochea} G. R. Bengochea  and R. Ferraro,   \href{http://dx.doi.org/10.1103/PhysRevD.79.124019}{Phys. Rev. D \textbf{79}, 124019  (2009)}.

\bibitem{Linder} E. V. Linder, \href{http://dx.doi.org/10.1103/PhysRevD.81.127301}{Phys. Rev. D \textbf{81}, 127301 (2010)}.

\bibitem{Ferraro} Y. F. Cai, S. Capozziello, M.  D. Laurentis and E. N. Saridakis,  \href{http://dx.doi.org/10.1088/0034-4885/79/10/106901}{Rept. Prog. Phys. \textbf{79}, 106901 (2016)}.

\bibitem{Oikonomou} V. K. Oikonomou and E. N. Saridakis,  \href{http://dx.doi.org/10.1103/PhysRevD.94.124005}{Phys. Rev. D {\bf 94}, 124005 (2016)}.

\bibitem{Ferraro1} R. Ferraro and F. Fiorini, \href{http://dx.doi.org/10.1103/PhysRevD.75.084031}{Phys. Rev. D \textbf{75}, 084031 (2007)}.

\bibitem{Ferraro2}  R. Ferraro and F. Fiorini, \href{http://dx.doi.org/10.1103/PhysRevD.78.124019}{Phys. Rev. D \textbf{78}, 124019 (2008)}.

\bibitem{Wright1} S. Bahamonde and M. Wright,   \href{http://dx.doi.org/10.1103/PhysRevD.92.084034}{Phys. Rev. D \textbf{92}, 084034  (2015)} [Erratum:\href{http://dx.doi.org/10.1103/PhysRevD.93.109901}{Phys. Rev. D \textbf{93}, 109901 (2016)}].

\bibitem{Wright2} S. Bahamonde, C. G. B\"{o}hmer and M. Wright,  \href{http://dx.doi.org/10.1103/PhysRevD.92.104042}{Phys. Rev. D \textbf{92}, 104042 (2015)}.

\bibitem{Myrzakulov} R. Myrzakulov, \href{http://dx.doi.org/10.1140/epjc/s10052-012-2203-y}{Eur. Phys. J. C \textbf{72}, 2203 (2012)}.

\bibitem{Barvinsky} A. Q.  Barvinsky, A. Y. Kamenshchik  and A. A. Starobinsky,  \href{http://dx.doi.org/10.1088/1475-7516/2008/11/021}{J. Cosmol. Astropart. Phys.  \textbf{11}, 021 (2008)}.

\bibitem{Panotopoulos} G. Panotopoulos, \href{http://dx.doi.org/10.1103/PhysRevD.76.127302}{Phys. Rev. D \textbf{76}, 127302 (2007)}.

\bibitem{Campo} S. Campo, R. Herrera    and A. Toloza,  \href{http://dx.doi.org/10.1103/PhysRevD.79.083507}{Phys. Rev.  D \textbf{79}, 083507 (2009)}.

\bibitem{Geng2} C. Q. Geng, C. C. Lee, E. N. Saridakis  and Y. P. Wu, \href{http://dx.doi.org/10.1016/j.physletb.2011.09.082}{Phys. Lett. B \textbf{ 704}, 384 (2011)}.

\bibitem{Geng4} C. Q.  Geng, C. C. Lee  and E. N. Saridakis, \href{http://dx.doi.org/10.1088/1475-7516/2012/01/002}{J. Cosmol. Astropart. Phys. \textbf{01}, 002 (2012)}.

\bibitem{Xu} C. Xu, E. N. Saridakis and  G. Leon,  \href{http://dx.doi.org/10.1088/1475-7516/2012/07/005}{J. Cosmol. Astropart. Phys. {\bf 07} (2012) 005}.

\bibitem{Geng3}  C. Q. Geng and Y. P. Wu,  \href{http://dx.doi.org/10.1088/1475-7516/2013/04/033}{J. Cosmol. Astropart. Phys.  \textbf{04} (2013) 033}.

\bibitem{Sahni} V. Sahni and S. Habib,  \href{http://dx.doi.org/10.1103/PhysRevLett.81.1766}{Phys. Rev. Lett.  \textbf{81} (1998) 1766}.

\bibitem{Uzan} J. P. Uzan,  \href{http://dx.doi.org/10.1103/PhysRevD.59.123510}{ Phys. Rev. D \textbf{59}, 123510 (1999)}.

\bibitem{Bartolo} N. Bartolo  and M. Pietroni, \href{http://dx.doi.org/10.1103/PhysRevD.61.023518}{Phys. Rev. D \textbf{61},  023518 (2000)}.

\bibitem{Faraoni} V. Faraoni,   \href{http://dx.doi.org/10.1103/PhysRevD.62.023504}{ Phys. Rev. D \textbf{62},  023504 (2000)}.

\bibitem{Elizalde}  E. Elizalde, S.  Nojiri  and S. D.  Odintsov,  \href{http://dx.doi.org/10.1103/PhysRevD.70.043539}{ Phys. Rev. D \textbf{70},  043539 (2004)}.

\bibitem{Hrycyna} O. Hrycyna  and M. Szydlowski, \href{http://dx.doi.org/10.1088/1475-7516/2009/04/026}{J. Cosmol. Astropart. Phys. \textbf{04}, 026 (2009)}.

\bibitem{Hrycyna2} O. Hrycyna  and M. Szydlowski,  \href{http://dx.doi.org/10.1103/PhysRevD.76.123510}{Phys. Rev. D \textbf{76}, 123510  (2007)}.

\bibitem{multifield}  V. Vardanyan  and L.  Amendola,  \href{http://dx.doi.org/10.1103/PhysRevD.92.024009}{Phys. Rev. D \textbf{92}, 024009 (2015)}.

\bibitem{Abedi} H.  Abedi  and A. M.  Abbassi,  \href{http://dx.doi.org/10.1088/1475-7516/2015/05/026}{J. Cosmol. Astropart. Phys. \textbf{05}, 026 (2015)}.

\bibitem{Kaiser} D. I. Kaiser,  \href{http://dx.doi.org/10.1103/PhysRevD.81.084044}{Phys. Rev. D \textbf{81}, 084044 (2010)}.

\bibitem{Yang} R. J. Yang,  \href{http://dx.doi.org/10.1209/0295-5075/93/60001}{Europhys. Lett. \textbf{93}, 60001 (2011)}.

\bibitem{Wright3} M. Wright,  \href{http://dx.doi.org/10.1103/PhysRevD.93.103002}{Phys. Rev. D \textbf{93}, 103002 (2016)}.


\bibitem{Li2} M. Li, R. X. Miao  and Y. J. Miao,  \href{http://dx.doi.org/10.1007/JHEP07(2011)108}{J. High Energy Phys.  \textbf{1107}, 108 (2011)}.

\bibitem{Ellis} G. F. R. Ellis, R. Maartens, M. A. H. MacCallum, Relativistic Cosmology, Cambridge University Press (2012).

\bibitem{Ray} S. Ray and U. Mukhopadhyay,  \href{http://dx.doi.org/10.1142/S0218271807011097}{Int. J. Mod. Phys. D \textbf{16}, 1791 (2007)}.

\bibitem{Uzan2} J. P. Uzan,   \href{http://dx.doi.org/10.12942/lrr-2011-2}{Living Rev. Rel. \textbf{14}, 2 (2011)}.

\bibitem{Dent} J. B. Dent, S. Dutta and E. N. Saridakis,  \href{http://dx.doi.org/10.1088/1475-7516/2011/01/009}{J. Cosmol. Astropart. Phys. {\bf 01}, 009 (2011)}.

\bibitem{Chen} S. H. Chen, J. B. Dent, S. Dutta and E. N. Saridakis,  \href{http://dx.doi.org/10.1103/PhysRevD.83.023508}{Phys. Rev. D {\bf 83}, 023508 (2011)}.

\bibitem{Zheng} R. Zheng and Q. G. Huang,  \href{http://dx.doi.org/10.1088/1475-7516/2011/03/002}{J. Cosmol. Astropart. Phys.  \textbf{03}, 002 (2011)}.

\bibitem{Wu1} Y. P.  Wu   and C. Q.  Geng,  \href{http://dx.doi.org/10.1007/JHEP11(2012)142}{J. High Energy Phys. \textbf{1211}, 142 (2011)}.

\bibitem{Nesseris} S. Nesseris, S. Basilakos and E.N. Saridakis,  \href{http://dx.doi.org/10.1103/PhysRevD.88.103010}{Phys. Rev. D {\bf 88},  103010 (2013)}.

\bibitem{Brans} C. Brans and R. H. Dicke, \href{http://dx.doi.org/10.1103/PhysRev.124.925}{Phys. Rev. {\bf 124}, 925 (1961)}.

\bibitem{Abedi2} H. Abedi and M. Salti, \href{http://dx.doi.org/10.1007/s10714-015-1935-z}{Gen. Rel. Grav. {\bf 47}, 93 (2015)}.

\bibitem{Barrow}
B. Li, T. P. Sotiriou and J. D. Barrow, \href{http://dx.doi.org/10.1103/PhysRevD.83.104017}{Phys. Rev. D \textbf{83}, 104017 (2011)}.

\bibitem{chameleon}
J. Khoury and A. Weltman,  \href{https://doi.org/10.1103/PhysRevD.69.044026}{Phys. Rev. D \textbf{69}, 044026 (2004)}.

\bibitem{symmetron}
K. Hinterbichler and J. Khoury, \href{https://doi.org/10.1103/PhysRevLett.104.231301}{Phys. Rev. Lett. \textbf{104}, 231301 (2010)}.

\bibitem{Boisseau} B. Boisseau, G. E. Farese, D. Polarski and A. A. Starobinsky, \href{http://dx.doi.org/10.1103/PhysRevLett.85.2236}{Phys. Rev. Lett. {\bf 85}, 2236 (2000)}.


\bibitem{Gu12}
J. A. Gu, C. C. Lee and C. Q. Geng, \href{https://doi.org/10.1016/j.physletb.2012.11.053}{Phys. Lett. B \textbf{718}, 722 (2013)}.

\bibitem{Jimenez02}
R. Jimenez, A. Loeb, \href{https://doi.org/10.1086/340549}{Astrophys. J., \textbf{573}, 37 (2002)}.


\bibitem{Fu11}
X. Fu, P. Wu and H. Yu, \href{https://doi.org/10.1142/S0218271811019372}{Int. J. Mod. Phys. D \textbf{20}, 1301 (2011)}.

\bibitem{Nesseris17}
S. Nesseris, G. Pantazis and L. Perivolaropoulos, \href{https://doi.org/10.1103/PhysRevD.96.023542}{Phys. Rev. D \textbf{96}, 023542 (2017)}.


\bibitem{Kofinas} G. Kofinas,  G. Leon and E. N. Saridakis, \href{http://dx.doi.org/10.1088/0264-9381/31/17/175011}{Class. Quant. Grav. \textbf{31}, 175011 (2014)}.

\bibitem{Kofinas1} G. Kofinas, and  E. N. Saridakis, \href{http://dx.doi.org/10.1103/PhysRevD.90.084044}{Phys. Rev. D \textbf{90}, 084044 (2014)}.

\bibitem{Bahamonde5} S. Bahamonde and C. G. B\"{o}hmer, \href{http://dx.doi.org/10.1140/epjc/s10052-016-4419-8}{Eur. Phys. J. C \textbf{76}, 578 (2016)}.

\bibitem{galileon1}
A. Nicolis, R. Rattazzi and E. Trincherini, \href{https://doi.org/10.1103/PhysRevD.79.064036}{Phys. Rev. D \textbf{79}, 064036 (2009)}.

\bibitem{galileon2}
A. De Felice and S. Tsujikawa, \href{https://doi.org/10.1103/PhysRevLett.105.111301}{Phys. Rev. Lett. \textbf{105}, 111301 (2010)}.

\bibitem{proca}
A. De Felice, L. Heisenberg, R. Kase, S. Mukohyama, S. Tsujikawa, Y. L. Zhang,  \href{https://doi.org/10.1088/1475-7516/2016/06/048}{J. Cosmol. Astropart. Phys, \textbf{1606}, 048 (2016)}.


\bibitem{Zhang14}
C. Zhang et al., \href{https://doi.org/10.1088/1674-4527/14/10/002}{Res. Astron. Astrophys., \textbf{14}, 1221 (2014)}.

\bibitem{Simon05}
J. Simon, L. Verde, R. Jimenez, \href{https://doi.org/10.1103/PhysRevD.71.123001}{Phys. Rev. D, \textbf{71}, 123001 (2005)}.

\bibitem{Moresco12}
M. Moresco et al., \href{https://doi.org/10.1088/1475-7516/2012/08/006}{J. Cosmol. Astropart. Phys., \textbf{8}, 006 (2012)}.

\bibitem{Chuang12}
C.-H. Chuang et al., \href{https://doi.org/10.1111/j.1365-2966.2012.21565.x}{Mon. Not. R. Astron. Soc., \textbf{426}, 226 (2012)}.

\bibitem{Moresco16}
M. Moresco et al., \href{https://doi.org/10.1088/1475-7516/2016/05/014}{J. Cosmol. Astropart. Phys., \textbf{05}, 014 (2016)}.

\bibitem{Stern10}
D. Stern, R. Jimenez, L. Verde, S. A. Stanford, M. Kamionkowski, \href{https://doi.org/10.1088/0067-0049/188/1/280}{Astrophys.J.Suppl., \textbf{188}, 280 (2010)}.

\bibitem{Moresco15}
M. Moresco, \href{https://doi.org/10.1093/mnrasl/slv037}{Mon. Not. R. Astron. Soc, \textbf{450}, L16 (2015)}.

\bibitem{Huterer17}
D. Huterer, D. L. Shafer, D. Scolnic and F. Schmidt,  \href{https://doi.org/10.1088/1475-7516/2017/05/015}{J. Cosmol. Astropart. Phys.  \textbf{05}, 015 (2017)}.

\bibitem{Turnbull12}
S. J. Turnbull et al., \href{https://doi.org/10.1111/j.1365-2966.2011.20050.x}{Mon. Not. Roy. Astron. Soc. \textbf{420}, 447 (2012)}.

\bibitem{Hudson12}
M. J. Hudson and S. J. Turnbull, \href{https://doi.org/10.1088/2041-8205/751/2/L30}{Astrophys. J. \textbf{751}, L30 (2013)}.

\bibitem{Davis11}
M. Davis et al., \href{https://doi.org/ 10.1111/j.1365-2966.2011.18362.x}{Mon. Not. Roy. Astron. Soc. \textbf{413}, 2906 (2011)}.

\bibitem{Feix15}
M. Feix, A. Nusser and E. Branchini, \href{https://doi.org/10.1103/PhysRevLett.115.011301}{Phys. Rev. Lett. \textbf{115}, 011301 (2015)}.

\bibitem{Song09}
Y. S. Song and W. J. Percival,  \href{https://doi.org/10.1088/1475-7516/2009/10/004}{J. Cosmol. Astropart. Phys. \textbf{0910}, 004 (2009)}.

\bibitem{Blake13}
C. Blake et al.,  \href{https://doi.org/10.1093/mnras/stt1791}{Mon. Not. Roy. Astron. Soc. \textbf{436}, 3089 (2013)}.

\bibitem{Howlett15}
C. Howlett et al., \href{https://doi.org/10.1093/mnras/stu2693}{Mon. Not. Roy. Astron. Soc. \textbf{449}, 848 (2015)}.

\bibitem{Samushia12}
L. Samushia, W. J. Percival and A. Raccanelli, \href{https://doi.org/10.1111/j.1365-2966.2011.20169.x}{Mon. Not. Roy. Astron. Soc. \textbf{420}, 2102 (2012)}.

\bibitem{Sanchez14}
A. G. Sánchez et al., \href{https://doi.org/10.1093/mnras/stu342}{Mon. Not. Roy. Astron. Soc. \textbf{440}, 2692 (2014)}.

\bibitem{Chuang16}
C. H. Chuang et al., \href{https://doi.org/10.1093/mnras/stw1535}{Mon. Not. Roy. Astron. Soc. \textbf{461}, 3781 (2016)}.

\bibitem{Blake12}
C. Blake et al., \href{https://doi.org/10.1111/j.1365-2966.2012.21473.x}{Mon. Not. Roy. Astron. Soc. \textbf{425}, 405 (2012)}.

\bibitem{Pezzotta16}
A. Pezzotta et al., \href{https://doi.org/10.1051/0004-6361/201630295}{Astron. Astrophys. \textbf{604}, A33 (2017)}.

\bibitem{Okumura16}
T. Okumura et al., \href{https://doi.org/10.1093/pasj/psw029}{Publ. Astr. Soc. Jap. {\bf 68}, 38 (2016)}.


\end{thebibliography}
\end{document}